\documentclass[%
 reprint,
 amsmath,amssymb,
 prl,
]{revtex4-2}

\usepackage[english]{babel}

\usepackage{graphicx}
\usepackage{dcolumn}
\usepackage{bm}
\usepackage{hyperref}
\usepackage[dvipsnames]{xcolor}
\usepackage{multirow}

\newcommand{\rvec}{ \mathbf{r} }

\newcommand{\fvec}{ \mathbf{f} }

\newcommand{\nhat}{ \mathbf{\hat{n}} }

\begin{document}

\preprint{APS/123-QED}

\title{Activity-driven emulsification of phase-separating binary mixtures}%

\author{Javier D\'iaz}
\affiliation{
Departament de F\'isica de la Mat\`eria Condensada, Universitat de Barcelona, Mart\'i i Franqu\'es 1, 08028 Barcelona, Spain 
}%
\affiliation{  
Universitat de Barcelona Institute of Complex Systems (UBICS), Universitat de Barcelona, 08028 Barcelona, Spain
}%
\author{Ignacio Pagonabarraga}%
 \email{ipagonabarraga@ub.edu}
\affiliation{
Departament de F\'isica de la Mat\`eria Condensada, Universitat de Barcelona, Mart\'i i Franqu\'es 1, 08028 Barcelona, Spain 
}%
\affiliation{  
Universitat de Barcelona Institute of Complex Systems (UBICS), Universitat de Barcelona, 08028 Barcelona, Spain
}%

\date{\today}

\begin{abstract}
Systems containing active components are intrinsically out of equilibrium, while binary mixtures reach their equilibrium configuration when complete phase separation is achieved. 
Active particles are found to stabilise non-equilibrium morphologies in phase separating binary mixtures by arresting coarsening by exerting active pressure that competes with surface tension driving forces. 
For moderate activities, an emulsion morphology is stabilised, where the droplet size is well-defined and controlled by activity. 
Conversely, the ability of active particles to drive phase-separated mixtures away from their equilibrium configuration is shown. 
A rich co-assembly behaviour is shown due to the competing energy scales involved in the system. 
\end{abstract}

\maketitle

\textbf{Introduction. }
Emulsification of phase-separating binary mixtures (BM) is relevant for several industrial applications where a precise control of the characteristic droplet size is desirable\cite{ferrari_molecular_2022}. 
The coarsening behaviour of pure BM phase-separating via Ostwald ripening is well understood in the LSW theory\cite{lifshitz_kinetics_1961,wagner_theorie_1961,konig_two-dimensional_2021} predicting a power law scaling of phase-separated domains $R(t)\sim t^{\alpha}$, while the inclusion of fillers has shown the ability to modify the coarsening behaviour\cite{balazs_multi-scale_2000}, or even arrest it altogether, such as in the case of Pickering emulsions\cite{jansen_bijels_2011} and bijels\cite{stratford_colloidal_2005}. 
Beyond BMs, two-phase coarsening governed by the LSW theory include motility-induced phase separation\cite{gonnella_motility-induced_2015,caporusso_dynamics_2022} or  planet formation\cite{tremaine_origin_2003}.  

Active particles (APs) are intrinsically out-of-equilibrium at the level of each particle, which consumes energy from the embedding medium to perform work, typically, self-propulsion. 
This has been shown to lead to a rich self-assembly behaviour, exhibiting phase separation\cite{tailleur_statistical_2008,cates_motility-induced_2015,digregorio_full_2018,hagen_brownian_2011} and polar order\cite{vicsek_novel_1995,sese-sansa_velocity_2018}. 
APs  are often found at fluid-fluid interfaces\cite{wang_enhanced_2015,wang_wetting_2016}. 
Furthermore, living active matter, such as bacteria, are often dispersed within a complex fluid composed of several species. 
APs are increasingly used for biological systems, in the presence of heterogeneous media. 

The out-of-equilibrium nature of systems containing active elements suggests the possibility of reaching non-equilibrium steady states different from the equilibrium counterparts. 
Non-equilibrium shape fluctuations have been encountered in droplets containing APs\cite{kokot_spontaneous_2022}, where collective motion under confinement was observed. 
Similarly, a deformable, soft confining medium can drive the spontaneous emergence of chiral collective motion in active filaments \cite{peterson_vesicle_2021}. 
The emergence of collective motion in systems involving APs under confinement has been reported for living\cite{wioland_confinement_2013} and inert particles\cite{scholz_surfactants_2021}. 
Conversely, droplets containing APs can lead to droplet self-propulsion\cite{gao_self-driven_2017}. 
Furthermore, the ability to introduce space-dependent active velocity has led to emergent anomalous diffusivity\cite{fernandez-rodriguez_feedback-controlled_2020}. 
The ability of APs to impact the equilibrium configuration of droplets motivates the study of their effect on the coarsening behaviour of passive phase-separating BMs.

\textbf{Model.} 
We use a mesoscopic model for a system containing $N_p$ APs with diameter $\sigma$ in a BM, described by the differences in concentration of A and B species $\psi(\rvec,t)=\phi_A(\rvec,t)-\phi_B(\rvec,t)$. 
The total free energy of the system has three contributions $F=F_{BM}+F_{cpl}+F_{pp}$, where $F_{BM}=\int d\rvec \left[ -1/2 \tau \psi^2+1/4 u\psi^4 +1/2 D (\nabla\psi)^2 \right]$ is a standard Ginzburg-Landau free energy, leading to a surface tension\cite{ohta_equilibrium_1986} of the BM $\gamma_{AB}=(2\sqrt{3}/2)\sqrt{D\tau^3}/u$. 
The particle-particle interaction free energy $F_{pp}$ is pairwise additive repulsive and penalises particle overlapping. 
The coupling free energy is 
\begin{equation}
    F_{cpl} = \sum_{i=1,N_p} c \int d\rvec \psi_c(r) \left[ \psi-\psi_0  \right]^2
    \label{eq:cpl}
\end{equation}
where $c$ specifies the scale of the particle-field interaction, the affinity parameter $\psi_0$ specifies the selectivity of the BM towards the particle and $\psi_c$ is a tagged function that determines the size of the particle.  
A characteristic energy scale can be extracted as $\epsilon_{cpl}=c\sigma^2 \psi_{eq}^2$ where $\psi_{eq}=\sqrt{\tau/u}$ is the equilibrium value of the order parameter following minimisation of the local terms in the BM free energy.

The state of the system is given by $\psi(\rvec,t)$, the position of APs $\rvec_i$ and their orientation $\varphi_i$, with the dynamics of the system controlled by three coupled equations
\begin{subequations}
\begin{equation}
    \frac{\partial \psi}{\partial t} = 
    M \nabla^2 \left( \frac{\delta F}{\delta \psi} \right)
    +\eta_{BM}(\rvec,t)
    \label{eq:Cahn}
\end{equation}
\begin{equation}
    \frac{\partial \rvec_i}{\partial t} = 
    v_a \nhat_i + \fvec/\gamma_t + 
    \sqrt{2D_t} \xi_{t}
    \label{eq:brown.t}
\end{equation}
\begin{equation}
    \frac{\partial \varphi_i}{\partial t} = 
    \sqrt{2D_r} \xi_{r}
    \label{eq:brown.r}
\end{equation}
\label{eq:dynamic}
\end{subequations}
where equation \ref{eq:Cahn} is the standard Cahn-Hilliard-Cook\cite{cahn_free_1958,cahn_free_1959,cahn_free_1959-1,cook_brownian_1970} equation for the dynamics of diffusive phase-separating mixtures. 
The random fluctuations term $\eta_{BM}$ satisfies fluctuation-dissipation theorem \cite{ball_spinodal_1990}. 
Equations \ref{eq:brown.t} and \ref{eq:brown.r} constitute the Active Brownian Particle model with the two being coupled by the unit vector $\nhat_i=(\cos\varphi_i,\sin\varphi_i)$ that dictates the direction of self-propulsion. 
The force acting on particles are of two origins: repulsive particle-particle forces $\fvec_{pp}=-\nabla F_{pp}$ and coupling forces $\fvec_{cpl}=-\nabla F_{cpl}$ due to the embedding field. 
The active velocity is given by $v_a$ and defines a swimming time scale $t_{s}=\sigma/v_a$. 
Furthermore, the rotational diffusive time scales is  $t_{rot}=D_{r}^{-1}$. 
The Einstein relation applies for each diffusive constant $D_t=k_BT/\gamma_t$ and $D_r=k_BT/\gamma_r$. 

The dimensionless Peclet number can be defined $Pe=t_{rot}/t_{s} \propto v_a$, characterising the persistence of the active motion.
Alternatively, it can be defined as the ratio of the active energy $\epsilon_a=v_a\gamma_t\sigma$ and the thermal scale $Pe=\epsilon_a/k_BT$.
We will use $\sigma$, $t_{rot}$ and $k_BT$ as units of length, time and energy, respectively, while $\psi$ is expressed in units of the equilibrium value $\psi_{eq}$. 
We consider a 2D system with size $L_x=L_y\approx 76$ and periodic boundary conditions leading to a AP surface fraction $\phi_p=N_p\pi \sigma^2/(4L_xL_y)$. 
A standard cell dynamic simulation\cite{oono_computationally_1987} scheme coupled with Brownian dynamics\cite{ginzburg_modeling_2000,diaz_hybrid_2022} is used to numerically solve Eq. \ref{eq:dynamic}. 
A full description of the model, as well as complete list of the parameters used can be found in the Appendix. 

\begin{figure}[ht]
    \centering
    \includegraphics[width=1.0\linewidth]{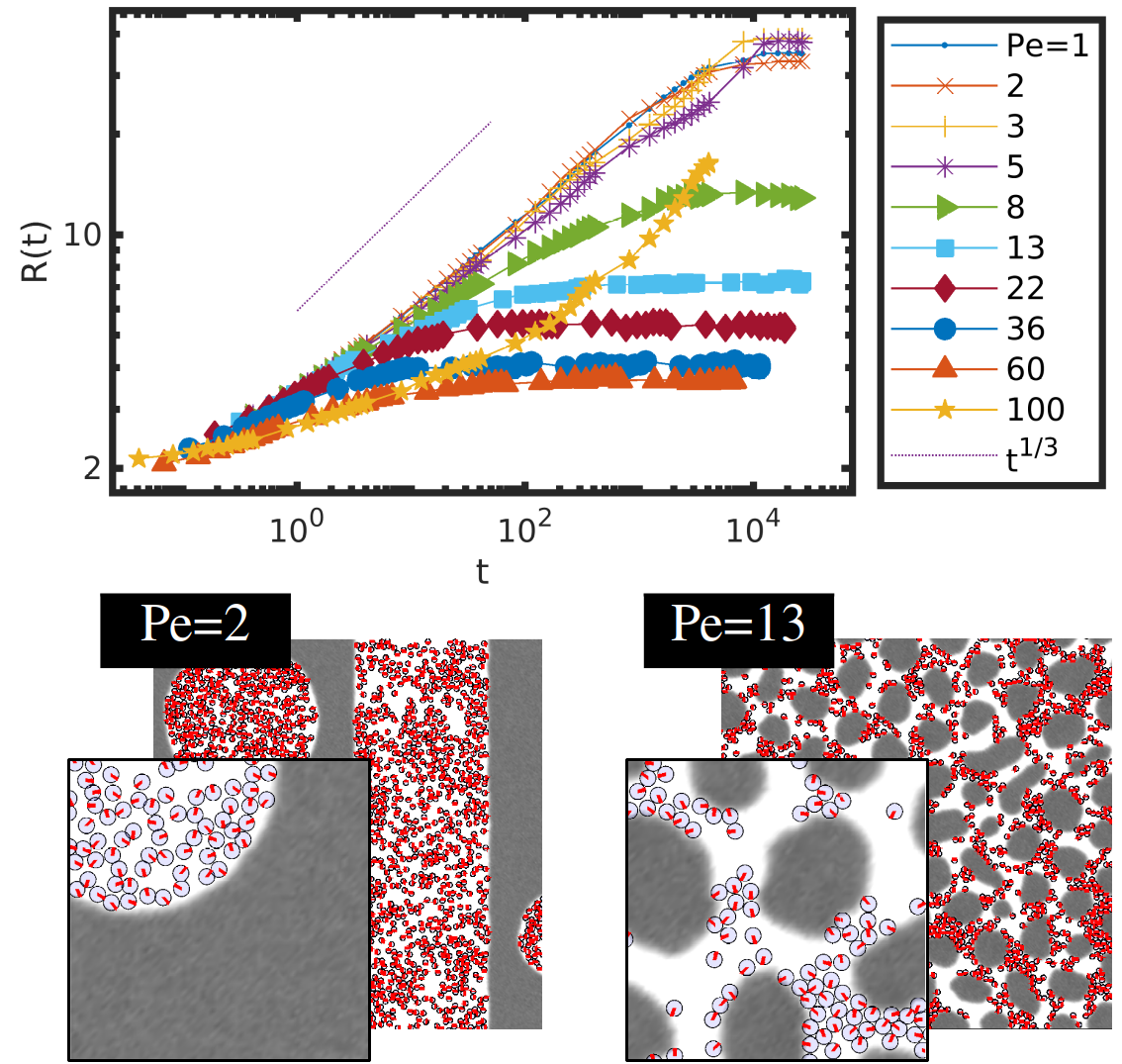}
    \caption{\textbf{Effect of APs on a phase separating mixture.}
    The characteristic length scale of the BM is shown in time along with the ideal $R(t)\sim t^{1/3}$ scaling, for different activity rates $Pe$. 
    In the bottom two snapshots are shown, corresponding to each of the regimes, with the orientation of each APs displayed as a red arrow. 
    }
    \label{fig:fromdis}
\end{figure}

\textbf{Results}
We consider a modest concentration of APs $\phi_p=0.2$ on a symmetric phase-separating mixture $\langle \psi \rangle=0$. 
The system is initialised from a random configuration of APs in a homogeneous distribution for the BM field $\psi$, modelling a quench from a disordered state. 
In the passive limit ($Pe\to 0$) APs are energetically favoured to segregate within the white domains in Fig. \ref{fig:fromdis} with $\psi_0=1$. 
In Fig. \ref{fig:fromdis}  we monitor the characteristic length scale of the binary mixture in time $R(t)$ calculated \textit{via} the scattering intensity of $\psi$ (see Eq. S22).

For small $Pe\lesssim 5$, the BM undergoes coarsening exhibiting a monotonic growth of $R(t)\sim t^{\alpha}$ over time, until complete phase separation is achieved in a long time scale. 
In the snapshot for $Pe=2$ a large percolating domain is formed, with a flat interface, while smaller droplets will eventually coalesce and macrophase separation is completed. 
The effect of APs on the exponent $\alpha$ will be subject of a future work.

However, for higher activity $Pe\gtrsim 8$ the coarsening of the BM appears to be arrested, with a plateau reached in the droplet size $R(t)\to R^*$ for late times, as shown in Fig. \ref{fig:fromdis}. 
The snapshot corresponding to $Pe=13$ in Fig. \ref{fig:fromdis} shows APs in the vicinity  the domain walls and preferentially pointing into the interface, which suggests the interplay between the active pressure and the surface-tension-driven coarsening of the interfaces. 
The snapshot for  $Pe=13$ clearly shows that the morphology of the mixture has changed from white droplets in a gray matrix, to gray droplets in a white matrix, stabilised by the active pressure exerted by the particles.
We note that, within this regime, as the activity increases, the stabilised droplet size $R^*$ decreases. 

Finally, for higher activity $Pe\gtrsim 60$, the BM is shown to recover its continuous growth and coarsening is resumed. 
We hypothesise that, for such high activity, APs carry a high enough active energy $\epsilon_a$ compared with the characteristic wetting energy of the BM, so that 
the dynamics of the APs and BM appear to be decoupled. 
Therefore, the coarsening behaviour recovers its expected scaling $R(t)\sim t^{1/3}$. 
However, the coarsening curve is not identical to the passive case, which motivates a closer study of the phase separation for high activity. 

The apparent coarsening prevention, shown in Fig. \ref{fig:fromdis} can be explained as APs extert active pressure on the BM interfaces. 
More broadly speaking, activity due to particles prevents the BM from reaching equilibrium and, instead, stabilise non-equilibrium morphologies of the mixture. 
In order to better support these claims, we consider an initially equilibrated flat interface separating A-rich and B-rich domains, with all APs initially located within the white phase with $\phi_p=0.2$. 
This is computationally advantageous to avoid the slow time scales associated with the macrophase separation and have better control over the geometry of the system. 
Therefore, at $Pe=0$ the initial condition is also the equilibrium configuration of the system.
By doing so, we directly assess the effect of APs on the BM equilibrium morphology, \textit{i.e.}, how activity can drive the system away from equilibrium. 
\begin{figure*}
    \centering
    \includegraphics[width=1.0\linewidth]{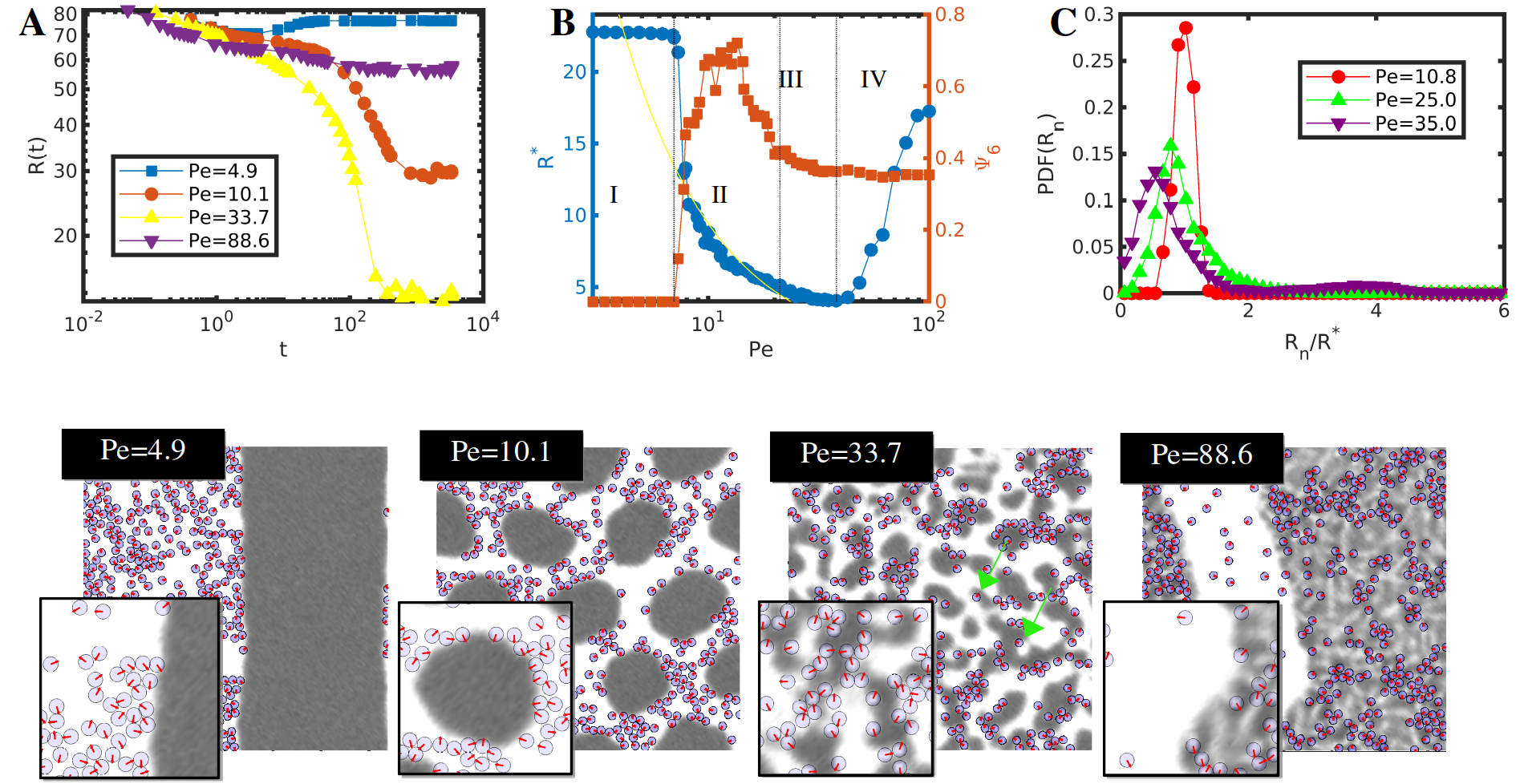}
    \caption{
    \textbf{Effect of APs on an equilibrated BM} with $\phi_p=0.2$. 
    In \textbf{A} the time evolution of the characteristic length scale of the BM is shown for selected values of $Pe$ for each of the regimes. 
    In \textbf{B} the steady state values of $R^*=R(t\to \infty)$ and the hexatic order parameter for the droplets $\Psi_6$ are shown. 
    In \textbf{C} the probability distribution function of the droplet size is shown for representative values of $Pe$, scaled with the mean value $R^*$. 
    Vertical dashed lines indicate the regime boundaries for $Pe=7.1$, $21.1$ and $38.1$. 
    In the bottom four snapshots are shown, corresponding to each of the regimes. 
    }
    \label{fig:Rtsnaps}
\end{figure*}

\begin{figure}
    \centering
    \includegraphics[width=0.9\linewidth]{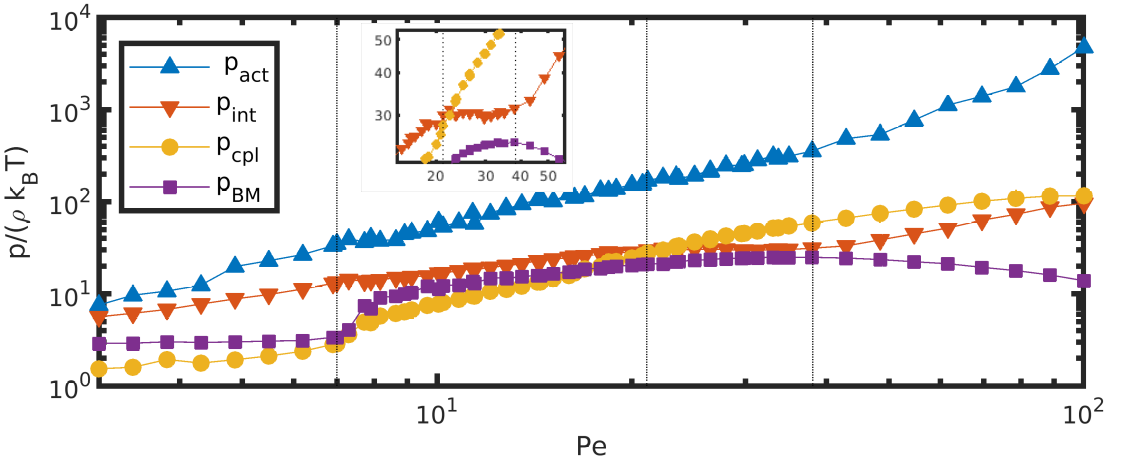}
    \caption{\textbf{Pressure profiles }in terms of $Pe$ for $\phi_p=0.2$, in the steady state, scaled with the ideal pressure $\rho k_BT$
    Different contributions to pressure are:
    $p_a$ active pressure, $p_{int}$ AP-AP interaction, $p_{cpl}$ coupling and BM pressure $p_{BM}$.  
    Vertical dashed lines indicate regime boundaries.
    The inset shows a detail of the behaviour in the unstable droplet phase (regime III). 
    }
    \label{fig:pressure}
\end{figure}

Fig. \ref{fig:Rtsnaps} shows the effect of APs on an equilibrated symmetric BM, quantified by the characteristic length scale $R(t)$ in \textbf{A} for selected values of $Pe$.
Again, for small $Pe$ $R(t)$ remains equal to the equilibrium size of the BM domain, which scales with the system size (see Fig. S8). 
However, for intermediate values of  $Pe$ a plateau is observed $R(t)\to R^*$, smaller than the system size, indicating that the steady-state morphology of the BM is not complete phase separation.  
In \textbf{B} the steady state behaviour is shown by $R^*$ in terms of $Pe$. 
Four different regimes can be identified visually, with the aid of the curve of $R^*$ and the hexatic order parameter $\Psi_6$, which characterises the global ordering of the BM domains into an hexagonal lattice. 
Furthermore, the pressure curves in Fig. \ref{fig:pressure} are used to identify critical points, shown as vertical bars in Fig. \ref{fig:Rtsnaps}. 

\begin{figure}
    \centering
    \includegraphics[width=0.99\linewidth]{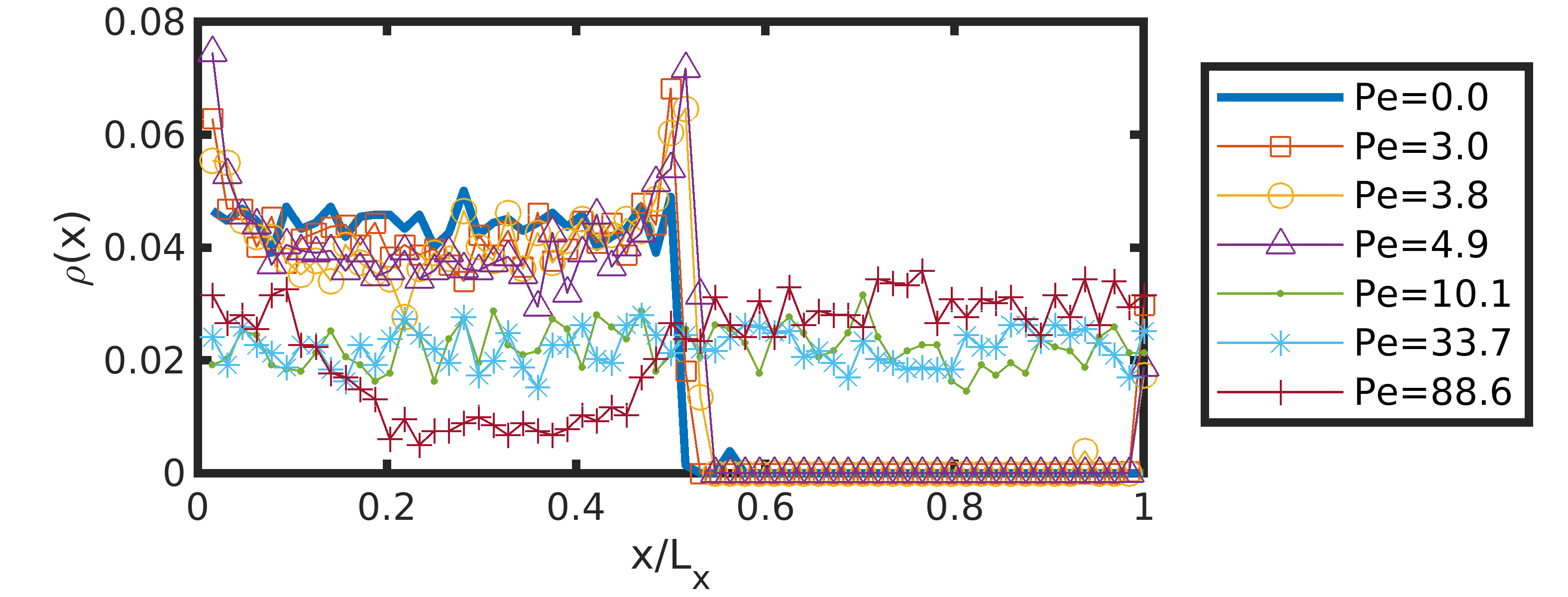}
    \caption{\textbf{Density profiles of APs} across the horizontal dimension of the system for selected values of $Pe$.
    Density is averaged over $10$ time steps after a steady state is reached.}
    \label{fig:density}
\end{figure}

In the low activity regime -regime I- APs do not significantly disturb the equilibrium morphology, with the system exhibiting a similar behaviour as confined APs within their preferred medium: 
APs accumulate at the soft confining walls, as shown in Fig. \ref{fig:density}, where density peaks appear close to the confining walls for $Pe<7.0$. 
In Fig. \ref{fig:pressure} the BM pressure is seen to be independent on the activity for $Pe<7$, indicating that the flat interface morphology remains unchanged. 

\begin{figure}
    \centering
    \includegraphics[width=1.0\linewidth]{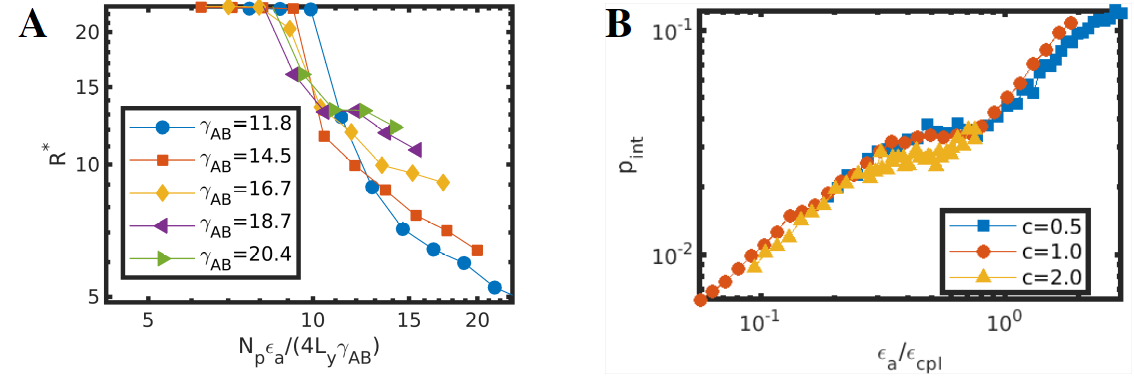}
    \caption{
    \textbf{Characterisation of transitions}.
    In \textbf{A} the characteristic domain size $R^*$ is shown for various values of BM interfacial tension $\gamma_{AB}$ in terms of the dimensionless energy ratio comparing total active energy $N_p\epsilon_a$ versus the interfacial energy. 
    In \textbf{B} the active energy per particle is compared with the coupling energy scale for various values of $c$. 
    }
    \label{fig:transition}
\end{figure}

For higher activity $7.0<Pe<21.1$, APs are able to drive the BM away from its equilibrium configuration and induce curvature -regime II- where an emulsion of droplets is stabilised due the APs. 
This is clearly marked as an increase in $p_{BM}$ in Fig. \ref{fig:pressure}, signaling the departure of the BM from its equilibrium configuration. 
The domain size distribution, shown in Fig. \ref{fig:Rtsnaps} \textbf{C}, indicates that a very well-defined droplet size $R^*$ can be identified. 
The onset of the I-II transition is the result of the competition between the total active energy for active particles near interfaces $\epsilon_a N_p$, compared with the total energy associated to the BM interface $\gamma_{AB}\Gamma \sim \gamma_{AB} 2L_y$, as shown in Fig. \ref{fig:transition} \textbf{A}, where the surface tension of the BM is shown to be the controlling parameter for the transition from regimes I and II, i.e., for the ability of APs' to drive the BM away from equilibrium. 
Moreover, the droplet morphology possesses hexatic order, quantified by a peak in $\Psi_6$ in Fig. \ref{fig:Rtsnaps}. 

We can estimate the scaling behaviour of $R^*(Pe)$ assuming a force balance of the total force due to surface tension $\gamma_{AB}$ on one droplet $f_{\gamma}\sim 2\pi R^* \gamma_{AB}$ and the total force due to activity exerted on a droplet $f_a^1\sim \gamma_t v_a N_p^1$ where $N_p^1=8\phi_p (R^*/\sigma)^2 $ is the average number of particles per droplet. 
Balancing these two forces acting on a single droplet we find $R^*/\sigma=\gamma_{AB}/(12k_BT Pe \phi_p)$. 
In Fig. \ref{fig:Rtsnaps} a yellow curve shows the $R^*\propto Pe^{-1}$ scaling which roughly predicts the behaviour of simulation data points.  
To support this mechanical basis for the emulsification of the BM, the role of the BM time scale is examined in Fig. S5, finding that the slower or faster BM time scale does not play a role in the emergence of the droplet morphology. 
Furthermore, the droplet morphology is irrespective of the initial conditions as shown in Fig. S6 which further shows the ability of APs to both arrest the phase separation of the BM and drive it out of equilibrium.  

For intermediate activity $21.1<Pe<38.1$, APs continue to drive the system away from equilibrium with the characteristic domain size $R^*$ being reduced with increasing $Pe$. 
However, in this regime APs do not stabilise droplets with an isotropic shape, instead, APs continue to push through the interfaces, resulting in isolated gray domains without a defined shape -regime III-. 
In this regime the droplets are not stabilised. 
This translates into a plateau in the interaction pressure in Fig. \ref{fig:pressure} as APs have more available real space to explore and AP-AP collisions are less likely to occur.
Furthermore, the droplet size distribution $PDF(R^*)$ in Fig. \ref{fig:Rtsnaps} \textbf{C} is considerably different to regime II: unstable droplets with amorphous shape have a much broader distribution of sizes compared to the $Pe\sim 10.1$ case. 
The onset of this transition is rationalised as the competition between the active energy per particle $\epsilon_a$ (as opposed to regime II where the total active energy is the controlling parameter), and the coupling energy $\epsilon_{cpl}$, which is associated to the surface tension of an AP immersed in the incompatible (gray) phase. 
This is clear in the collapsed of the interparticle pressure into a single curve for three different values of the AP-BM coupling parameter $c$, shown in Fig. \ref{fig:transition} \textbf{B} for varying $\epsilon_a/\epsilon_{cpl}$, which is reminiscent of the bond number for gravitational forces.

Finally, for large activities $Pe>38.1$ the APs active energy is considerably larger than the coupling energy due to the embedding BM.
Therefore, the dynamics of the APs and the BM appear to be largely decoupled -regime IV-, with the BM increasingly recovering its equilibrium morphology towards complete phase separation. 
On the onset of the transition, in Fig. \ref{fig:Rtsnaps} \textbf{C} for $Pe=35$, it is possible to see the emergence of large domain sizes, albeit with low probability. 
On the other hand, as activity grows, APs are increasingly unaffected by the forces originating from the BM, except for a slightly reduced effective mobility when dispersed within the incompatible phase. 
This translates into more accumulation of APs within the gray region, which can be seen visually in Fig. \ref{fig:Rtsnaps} for $Pe=88.6$ and quantified in the density profiles in \ref{fig:density}, in sharp contrast with the low activity regime for $Pe<7$.

\textbf{Conclusions}
APs have been shown to drive phase-separating mixtures away their equilibrium configuration due to the competition between the active energy carried by the colloids, and the surface tension driving phase separation. 
This is shown both in the ability of APs to  prevent the BM from reaching its equilibrium morphology (Fig. \ref{fig:fromdis}) as well as driving the system away from equilibrium (Fig. \ref{fig:Rtsnaps}). 
In both cases APs can stabilise intermediate states characterised by droplets, composed of the incompatible species to the APs, within a matrix of the soluble phase. 
These droplets have a well-defined size and, in a slow time scale, can form organised hexagonal structures. 
The condition for droplet formation is that the active pressure exerted by the APs is comparable with the surface tension of the BM. 
In this sense, APs are surfactant-like despite being, at equilibrium, completely dispersed within one phase. 
Contrary to Pickering emulsions, where a high surface coverage is required, the size of the droplets is controlled by the activity. 
The wetting of the APs additionally selects the curvature of the droplets and determines the species confined in droplets.  
On the other hand, a secondary transition is observed, where the active energy of each particle is considerably larger than the wetting forces of the AP. 
It is precisely in the intermediate active energy values, where the active forces are enough to deform interfaces but not large enough to penetrate into the interface, where the emulsion is stabilised.

\begin{acknowledgments}
J.D. acknowledges financial support from the Spanish Ministry of Universities through the Recovery, Transformation and Resilience Plan funded by the European Union (Next Generation EU), and Universitat de Barcelona.
I.P. acknowledges support from Ministerio de Ciencia, Innovaci\'on y
Universidades MCIU/AEI/FEDER for financial support under
grant agreement PID2021-126570NB-100 AEI/FEDER-EU, from
Generalitat de Catalunya  under Program Icrea Acad\`emia and project 2021SGR-673.
\end{acknowledgments}

\renewcommand{\thefigure}{S\arabic{figure}}
\setcounter{figure}{0}  
\renewcommand\theequation{S\arabic{equation}}
 \setcounter{equation}{0}   
\appendix

\section{Supporting information}

We present additional results and details to facilitate reproducibility. 

\section{Full model description}
For clarity sake, in this section we show a complete description of the model. 

We consider a hybrid system composed of $N_p$ APs with positions $\rvec_i$ and orientations given by $\phi_i$ for the $i$th particle. 
The BM is described by differences in concentration of species A and B \textit{via} the order parameter $\psi(\rvec,t)=\phi_A(\rvec,t)-\phi_B(\rvec,t)$. 

The thermodynamic state of the system is specified by the total free energy, which has three terms, 
\begin{equation}
    F = 
    F_{BM} + F_{cpl}+F_{pp}
\end{equation}
respectively, free energy of the BM, coupling term and particle-particle interaction term. 
The free energy is assumed to be expressed in units of $k_BT$.
The BM free energy is 
\begin{equation}
    F_{BM} = \int d\rvec \left[ 
    -\frac{1}{2} \tau \psi^2 +\frac{1}{4} u \psi^4
    +\frac{1}{2} D (\nabla \psi)^2
    \right]
    \label{eq:esi.BM}
\end{equation}
where $\tau$ is related to the Flory-Huggins of the BM, $u$ specifies the amplitude of the concentration fluctuations and $D$ controls the width of the interface. 
In fact, the amplitude of the fluctuations can be analytically derived from the local terms in Eq. \ref{eq:esi.BM} as $\psi_{eq}=\sqrt{\tau/u}$ which we refer as the equilibrium values of the order parameter, that is, the values that $\psi$ takes in the bulk region when demixing takes place.

The coupling free energy is
\begin{equation}
    F_{cpl} = \sum_{i=1,N_p} c \int d\rvec \psi_c(r) \left[ \psi-\psi_0  \right]^2
    \label{eq:esi.cpl}
\end{equation}
where $c$ specifies the scale of the particle-field interaction, $\psi_0$ specifies the selectivity of the BM towards the particle and $\psi_c$ is a tagged function that determines the size of the particle.  
We choose a functional form\cite{tanaka_simulation_2000}
\begin{equation}
    \psi_c(r<R)=\exp\left[1-\frac{1}{1-(r/R)^2} \right]  
\end{equation}
and $\psi_c(r\geq0)=0$ which has a compact form and vanishing derivative at the cutoff $r=R$. 
A characteristic coupling energy scale can be defined as $\epsilon_{cpl}=c\sigma^2\psi_{eq}^2$. 

The particle-particle free energy term is
\begin{equation}
    F_{pp} = 
    \sum_{i\neq j} U(r_{ij})
\end{equation}
where the pairwise additive repulsive potential is 
\begin{equation}
    U(r_{ij}) = U_0 exp(1-r_{ij}/\sigma)/(r_{ij}/\sigma)
\end{equation}
where $U_0$ specifies the energetic scale of the repulsive particle-particle interaction and $\sigma=2R$ is the diameter of the particle and also the cut-off of the interaction $U(r_{ij}>\sigma)=0$ where $r_{ij}$ is the distance between particle pairs. 

The three coupled dynamic equations are 
\begin{subequations}
\begin{equation}
    \frac{\partial \psi}{\partial t} = 
    M \nabla^2 \left( \frac{\delta F}{\delta \psi} \right)
    +\eta_{BM}(\rvec,t)
    \label{eq:esi.Cahn}
\end{equation}
\begin{equation}
    \frac{\partial \rvec_i}{\partial t} = 
    v_a \nhat_i + \fvec_i/\gamma_t + 
    \sqrt{2D_t} \xi_{t}(t)
    \label{eq:esi.brown.t}
\end{equation}
\begin{equation}
    \frac{\partial \varphi_i}{\partial t} = 
    \sqrt{2D_r} \xi_{r} (t)
    \label{eq:esi.brown.r}
\end{equation}
\label{eq:esi.dynamic}
\end{subequations}
with $\xi_t(t)$ and $\xi_r(t)$ are zero-mean uncorrelated noises. 
The translational and rotational diffusion coefficients are related to each other as 
$D_t=\sigma^2 D_r/3$ while Einstein relationship applies $D_t=k_BT/\gamma_t$ and $D_r=k_BT/\gamma_r$. 
The force acting on particle $i$ in Eq. \ref{eq:esi.brown.t} has two contributions $\fvec_i=\fvec_i^{pp}+\fvec_i^{cpl}$, respectively due to the particle-particle interaction 
$\fvec_i^{pp}=-\nabla_iF_{pp}$ and the coupling interaction $\fvec_i^{cpl}=-\nabla_iF_{cpl}$, where the derivative is with respect to the particle position $\nabla_i\equiv \partial/\partial \rvec_i$.

Eq.\ref{eq:esi.Cahn} can be made explicit as 
\begin{equation}
\begin{split}
    \frac{\partial \psi}{\partial t} = &
    M \nabla^2 
    \left[
    -\tau \psi+u\psi^3 -D\nabla^2\psi +2c\psi_c (\psi-\psi_0)
    \right] \\
    &+\eta_{BM}(\rvec,t)
\end{split} 
\end{equation}
where the random fluctuation term for the BM satisfies the fluctuation-dissipation theorem as\cite{ball_spinodal_1990}
\begin{equation}
    \langle \eta(\rvec,t) \eta(\rvec',t')\rangle =
-k_B T M \nabla^2 \delta(\rvec-\rvec')
\delta(t-t')
\end{equation}

In order to reduce the number of independent parameters and to extract the relevant dimensionless quantities, let's consider the dimensionless form of the dynamic equations.
Primed (') magnitudes are dimensionless. 
We consider the following changes 
\begin{subequations}
    \begin{equation}
        \rvec \to \rvec' \sigma
    \end{equation}
    \begin{equation}
        t\to t' t_{rot}
    \end{equation}
    \begin{equation}
        \psi\to \psi' \psi_{eq}
    \end{equation}
\end{subequations}
where we are expressing lengths in units of the AP diameter $\sigma$ and  times in units of the rotational diffusion time $t_{rot} = D_r^{-1}$. 
Importantly, the concentration fluctuation field is expressed in units of $\psi_{eq}=\sqrt{\tau/u}$. 
This equilibrium amplitude of the concentration field can be derived analytically from the local terms in the BM free energy density in Eq. \ref{eq:esi.BM} (minimisation of the first and second term with respect to $\psi$, assuming $\tau$ and $u$ positive). 

The noise term in the Cahn-Hilliard-Cook equation can be made dimensionless by defining 
\begin{equation}
    \xi_{BM}(\rvec,t)=\eta(\rvec,t)/\sqrt{2k_BT M} 
\end{equation}
and making the change 
\begin{equation}
    \xi_{BM}(\rvec,t) = \xi_{BM}'(\rvec,t)\sigma^{-1} t_{rot}^{-1/2}
\end{equation}
following the fluctuation-dissipation theorem expressions for the amplitude and correlations of the noise. 
Then, the coupled dynamic equations can be rewritten in dimensionless form as 
\begin{subequations}
    \begin{equation}
    \begin{split}
        \frac{\partial \psi'}{\partial t'} = 
        \frac{t_{rot}\tau M }{\sigma^2} (\nabla')^2 \\
        \left[ 
        -\psi'+\psi'^3-\frac{D}{\tau} \sigma^{-2} (\nabla')^2 \psi' + 2c'\psi_c(\psi'-\psi_0')
        \right]  \\
        + \frac{t_{rot}^{1/2}}{\psi_{eq}\sigma}\sqrt{2k_BT M} \xi_{BM}'(\rvec,t)
        \label{eq:esi.cahn.1}
    \end{split}
    \end{equation}
    \begin{equation}
        \frac{\partial \rvec_i'}{\partial t'}=
        Pe \nhat_i(t) + 
        \frac{1}{3}U_0'\fvec_i^{pp\prime}+
        \frac{1}{3}\epsilon_{cpl}'\fvec_i^{cpl\prime} +
        \sqrt{2/3} \xi_{t}'(t)
    \end{equation}
    \begin{equation}
        \frac{\partial \varphi_i}{\partial t'} =
        \sqrt{2} \xi_{r}'(t)
    \end{equation}
\end{subequations}
where the dimensionless form of the two coupled Brownian equations are standard, since we are using AP natural units as units of dimensions. 
The ratio between the diffusive rotational time scale of the AP and the AP swimming time is defined as the Peclet number 
\begin{equation}
    Pe = \frac{t_{rot}}{t_{s}}
\end{equation}
. 
Meanwhile, two dimensionless energetic parameters are defined as 
\begin{subequations}
    \begin{equation}
        U_0'=\frac{U_0}{k_BT}
    \end{equation}
    \begin{equation}
        \epsilon_{cpl}'=\frac{\epsilon_{cpl}}{k_BT}
    \end{equation}
\end{subequations}
.

The Cahn-Hilliard-Cook dynamic equation deserves a closer look where new parameters are defined. 
Parameter $D$ (prefactor of square gradient term) is well known to be related to the interface width in Ginzburg-Landau-like models, with $L_w=\sqrt{D/\tau}$. 
Due to the use of $\sigma$ as unit of length, we define a dimensionless parameter 
\begin{equation}
    \rho_w = L_w/\sigma 
\end{equation}

In Eq. \ref{eq:esi.cahn.1} there is a prefactor to the deterministic part of the equation, which specifies the time scale of the BM. 
Again, in the standard Cahn-Hilliard dynamics a diffusive time scale can be defined as $t_{BM}=L_w^2/(M\tau)$.  
We can then define 
\begin{equation}
    \tilde{\eta} =  \frac{t_{rot}\tau M }{\sigma^2} = \frac{t_{rot}}{L_w^2/(\tau M)} \left( \frac{L_w}{\sigma} \right)^2=
    \eta \rho_w^2
\end{equation}
where we can define a purely time ratio
\begin{equation}
    \eta = \frac{t_{rot}}{t_{BM}}
\end{equation}
as the ratio between the diffusive time scale associated with the rotational degrees of freedom of the AP, and the relaxation time of the BM. 

Additionally, a secondary Peclet number can be defined as 
\begin{equation}
    Pe' = 
    \frac{t_{BM}}{t_{s}} = 
    \eta^{-1} Pe
\end{equation}
which compares the diffusive time of the BM and the swimming time of the AP. 
We can see that there are three independent time scales in the system $t_{BM}$, $t_{rot}$ and $t_{s}$ where the first two are passive. 
Then, we can specify the dynamic properties of the system with two dimensionless time scales, for example, $Pe$ and $Pe'$. 

We can also compare the active energy to the coupling energy 
\begin{equation}
    \epsilon_{cpl}^a = 
    \frac{\epsilon_a}{\epsilon_{cpl}} = 
    3 Pe \left( \frac{k_BT}{\sigma^2 c}  \right)
\end{equation}

Finally, the coupling term can be made dimensionless as 
\begin{equation}
    c'=\frac{c}{\tau}
\end{equation}
and the rescaled affinity parameter is clearly 
\begin{equation}
    \psi_0' = \psi_0/\psi_{eq}
\end{equation}
indicating that an affinity $\psi_0'=-1$ leads to complete solubility within the negative phase of the BM.

Hereafter we drop the primed (') notation by assuming $\sigma$ and the unit of length, $t_{rot}$ for times and $\psi_{eq}$ for concentration fluctuations.
By doing so, the three dynamic equations can be written as
\begin{subequations}
    \begin{equation}
        \begin{split}
            \frac{\partial\psi}{\partial t}=
            \tilde{\eta}
            \nabla^2\left[
            -\psi+\psi^3-\rho_w^2\nabla^2\psi
            +2c\psi_c(\psi-\psi_0)
            \right]\\
            +\sqrt{2\epsilon_0\tilde{\eta}} \xi_{BM}
        \end{split}
    \end{equation}
    \begin{equation}
        \frac{\partial \rvec_i}{\partial t}=
        Pe \nhat_i(t) + 
        \frac{1}{3}U_0\fvec_i^{pp}+
        \frac{1}{3}\epsilon_{cpl}\fvec_i^{cpl} +
        \sqrt{2/3} \xi_{t}(t)
    \end{equation}
    \begin{equation}
        \frac{\partial \varphi_i}{\partial t} =
        \sqrt{2} \xi_{r}(t)
    \end{equation}
\end{subequations}

\section{Numerical scheme}

The numerical scheme for solving Eq. 2 is based on hybrid\cite{diaz_hybrid_2022} CDS/Brownian dynamics with forward Euler integration\cite{oono_computationally_1987,oono_study_1988}. 
In CDS the Laplacian is implemented as $\nabla^2 \psi = 4/(\delta x)^2\left[ \langle\langle \psi \rangle\rangle- \psi  \right] $
where the average operator included nearest-neighbor (NN) and next-nearest-neighbor (NNN) to improve the isotropy of the scheme, 
\begin{equation}
    \langle\langle \psi \rangle\rangle = 
    \frac{1}{6} \sum_{NN} \psi +
    \frac{1}{12} \sum_{NNN} \psi 
\end{equation}

The space discretisation is implemented such that the relevant length scales are larger than the grid spacing. 
In particular, the AP diameter is $\sigma/ \delta x=3.38$ in units of grid  spacing.

The time discretisation scheme takes into account the presence of several time scales in the dynamic model: 
time associated with the BM relaxation $t_{BM}$, times associated with the translational $t_{t}$ and rotational $t_{r}$ diffusion of the APs and the swimming time scale $t_s$ associated with the active motion. 
We consistently implement a discretisation $\tilde{\delta t}=\delta t/ t_{faster}=0.01225$ where $t_{faster}$ is the fastest (smallest) time scale present in the system, that is, the minimum out of all the described characteristic times.

\section{List of parameters}

This work involves a large number of parameters.
For clarity sake we provide all the values in dimensionless form as shown in the previous section. 
The main parameters that we explore are $Pe$ and $\phi_p$. 

The time scales of the system are set as $\eta=10$.
The relative length scale of the system is set to $\rho_w=1/4$, so that the AP diameter is larger than a typical interface.
We select a thermal scale $k_BT=0.05$. 
The BM parameters are standard in the literature\cite{pinna_large_2012}. 
In units of $k_B T$ are $\tau=7$, $u=10$ and $D=5.0$. 
The AP affinity is $\psi_0=+1$. 
The AP-BM interaction strength is $c'=2.9$. 
The particle-particle scale is $U_0=10$, however, we note that we increase this with $Pe$ (scaling with $Pe$) in order to prevent overlapping of two colliding AP with opposite velocities. 

\section{Observables}

List of quantities used to study the system 

\textbf{Characteristic lengthscale of the BM}
We calculate the characteristic length scale of the BM $R(t)$ by calculating first the scattering function of the order parameter 
\begin{equation}
    S(\textbf{q},t) = |\psi_{\textbf{q}}(t)|^2
\end{equation}
where $\psi(\textbf{q}) = F[\psi(\textbf{r})]$ is the Fourier transform of the order parameter. 
The radial scattering function $S(q,t)$ can be used to calculate the characteristic length scale  
\begin{equation}
    R(t) = 2\pi /q^*(t); q^*(t) = \langle qS(q,t) \rangle
    \label{eq:esi.rt}
\end{equation}
where $\langle *\rangle$ is the radial average. 
The quantity $R(t)$ can be used to study the coarsening of the BM. 
If a steady state is reached, we calculate $R^*=\langle R(t) \rangle$ with the time average taken after steady state is reached, over several time steps. 

\textbf{Domain analysis}
The characteristic length scale $R(t)$ is a global average that ignores the diversity in domain sizes that are frquent for collections of droplets. 
Domain analysis tools are used to identify the grid points belonging to a B-rich (or A-rich) domain  based on the order parameter field $\psi(\rvec,t)$, 
\begin{equation}
    \psi(\rvec_{-})<0 \rightarrow  \{ \rvec_{-} \}
    \label{eq:domainanalysis}
\end{equation}
Standard cluster analysis tools are used to automatically identify domains (\textit{i.e.}, clusters of connected interface nodes). 
From the coordinates of interfaces it is possible to obtain domain centroids, which will be used to calculate $\Psi_6$ (see below in Eq. \ref{eq:psi6}). 
Furthermore, it is possible to identify the radius of each domain $R_{n}$ to obtain the probability distribution function $PDF(R_n)$. 



\textbf{Hexatic order parameter $\Psi_6$}
We first perform domain analysis on the order parameter field $\psi(\rvec,t)$ to identify the centre of each BM domains. 
The coordinates of these centroids are $X_i,Y_i$ with $i$ being a BM isolated domain. 
The hexatic order parameter is then defined as 
\begin{equation}
    \Psi_6= \langle  e^{6i\theta_{ij}} \rangle
    \label{eq:psi6}
\end{equation}
where $\langle * \rangle$ represents an average over time and immediate neighboring particles. 
Neighboring particles are determined based on Voronoi tesselation analysis. 


\textbf{Pressure}
The ideal pressure for a system of non-interacting particles is $p_{id} = \rho k_BT$.  
We have four contributions for the total excess pressure of the system are
\begin{equation}
    p = p_{act} + p_{int}  + p_{cpl} + p_{BM}
\end{equation}
respectively for the active, interparticle, coupling and BM contributions. 
We use standard methods to calculate the pressure in all cases, based on the virial pressure. 
In all cases, the pressure is calculated as the trace of the stress tensor as $p_{k}=Tr(\sigma_{\alpha\beta}^k)$ where $k$ is each of the pressure contributions and $\alpha\beta$ are the  cartesian components of the tensor. 

For the BM pressure the contributions arise from the interfaces of the BM\cite{ohta_anomalous_1993}
\begin{equation}
    \sigma_{\alpha\beta}^{BM}= -\frac{V}{D} \int d\rvec \partial_{\alpha} \psi \partial_{\beta} \psi  
\end{equation}
where $\partial_{\alpha}$ is the spatial derivative in the $\alpha$ Cartesian coordinate, and $V$ is the volume (area in 2D) of the system. 

For the coupling pressure we have to consider that AP $i$ interacts with the embedding BM through the order parameter field $\psi(\rvec,t)$. 
For this reason, we use a virial-like pressure calculation where particle $i$ in continuous space $\rvec_i$ is considered to interact with fictitious particle $j$ corresponding to node point $\rvec_j$ corresponding to $\psi(\rvec_j)$. 
The coupling force contribution $ f_{ij}^{cpl,\alpha}$
\begin{equation}
    \sigma_{\alpha\beta}^{cpl}=
    \frac{1}{2V} \sum_{ij} f_{ij}^{cpl,\alpha}x_{ij}^{\beta}
\end{equation}
where $f_{ij}^{\alpha}$ is the $\alpha$ component of the force contribution arising from the interaction between particle $i$ and node point $\rvec_j$. 
This force arises from the gradient of the coupling free energy in Eq. 1. 
On the other hand, the distance $x^{\beta}_{ij}$ is the $\beta$ component of the distance between particle $i$ position (continuous) and node point $\rvec_j$, $x_{ij}^{\beta}=x_i^{\beta}-x_j^{\beta}$. 
Conceptually, we can understand that particle $i$ interacts with he collection of monomers at node $\rvec_j$. 

The interacting pressure $p_{int}$ arises from the repulsive forces between APs. 
We use the virial expression 
\begin{equation}
    \sigma_{\alpha\beta}^{int}=
    \frac{1}{2V} \sum_{ij} f_{ij}^{\alpha}x_{ij}^{\beta}
\end{equation}
where $i,j$ are particle pairs, $f_{ij}^{\alpha}$ is the repulsive force and $x_{ij}$  is the distance between particles. 

Finally, the active pressure is calculated as\cite{solon_pressure_2015,speck_ideal_2016}
\begin{equation}
    p_{a} = 
    \frac{v_a \gamma_t}{2V} \sum_i \langle \nhat_i \cdot \rvec_i \rangle
\end{equation}


\section{Additional simulations}

\subsection{Dilute limit}

\begin{figure}
    \centering
    \includegraphics[width=1.0\linewidth]{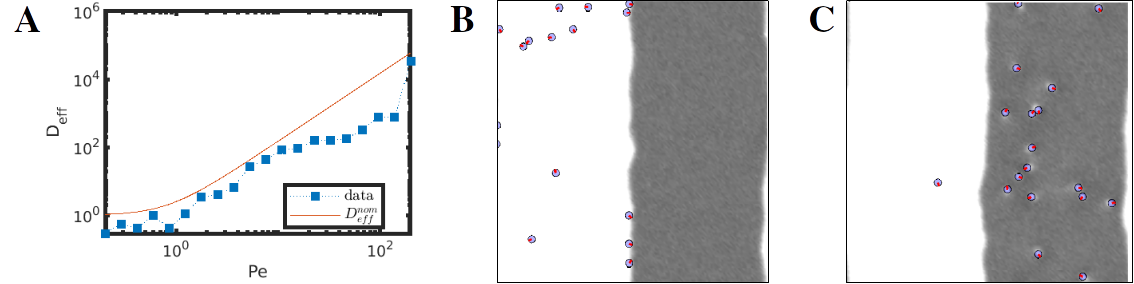}
    \caption{
    \textbf{Dilute regime} with $\phi_p=0.01$ initialised from equilibrium configuration. 
    In \textbf{A} the effective diffusion is shown along with the analytical value. 
    B and C show snapshots for representative values of $Pe=15.7$ and $Pe=67.2$, respectively. 
    }
    \label{fig:dilute}
\end{figure}

In the extremely dilute limit $\phi_p\to 0$, we can expect that APs do not significantly disturb the coarsening behaviour or the equilibrium morphology of the BM. 
Nonetheless, it is useful to study this limit to obtain insight on the dynamics of APs within BMs in a controlled way.
Fig. \ref{fig:dilute} \textbf{A} shows the effective  diffusion constant $D_{eff}$ for a small concentration $\phi_p=0.01$, which is numerically fitted from $MSD(t)$ curves in the long-time regime. 
The fitted data is compared with the analytical long-time behaviour\cite{bechinger_active_2016}
\begin{equation}
    D_{eff}^{a} = 
    1 +\frac{6}{4} Pe^2
\end{equation}
which shows that APs behave in a similar fashion as free APs when $Pe<20$, for higher activity the effective diffusivity changes its trend and its growth is slowed down. 
By looking at the snapshots in Fig. \ref{fig:dilute}, it is clear that APs with moderate activity remain at their preferred domain (white phase), while higher activity leads to the APs being able to penetrate into their incompatible gray phase. 
As APs enter into the gray phase, coupling forces slow down the instant speed of the APs, which leads to reduced diffusivity compared with the analytical value, as well as accumulation at the gray phase: 
for a given $Pe$, APs within the gray phase are slower and therefore spend longer time within the incompatible phase. 
For this reason, in $C$ APs are mostly find within the right side of the system. 

\begin{figure}
    \centering
    \includegraphics[width=1.0\linewidth]{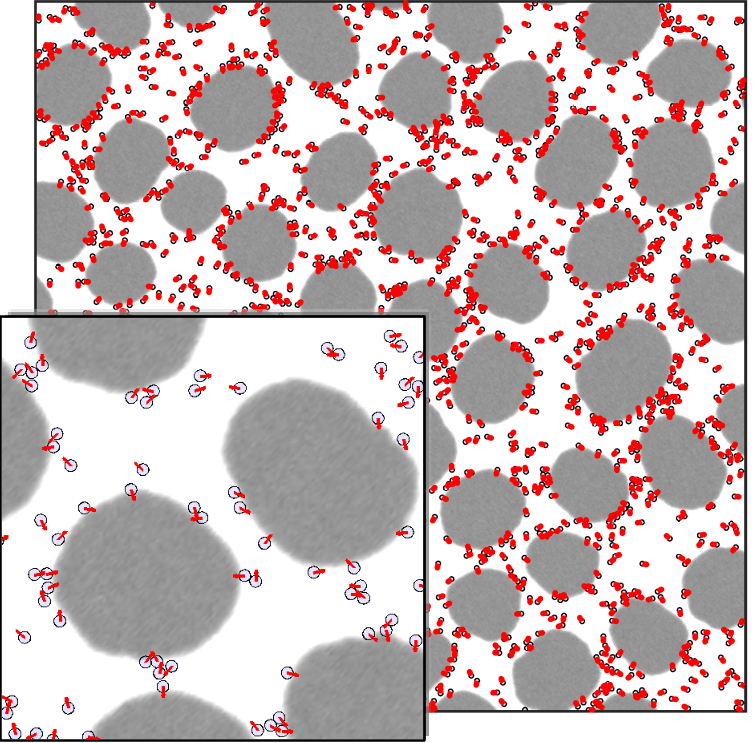}
    \caption{Large scale simulation of the same system as in Fig. 2 with $Pe=10$}
    \label{fig:hex.large}
\end{figure}

\subsection{Role of AP concentration on droplet morphology}

\begin{figure}
    \centering
    \includegraphics[width=1.0\linewidth]{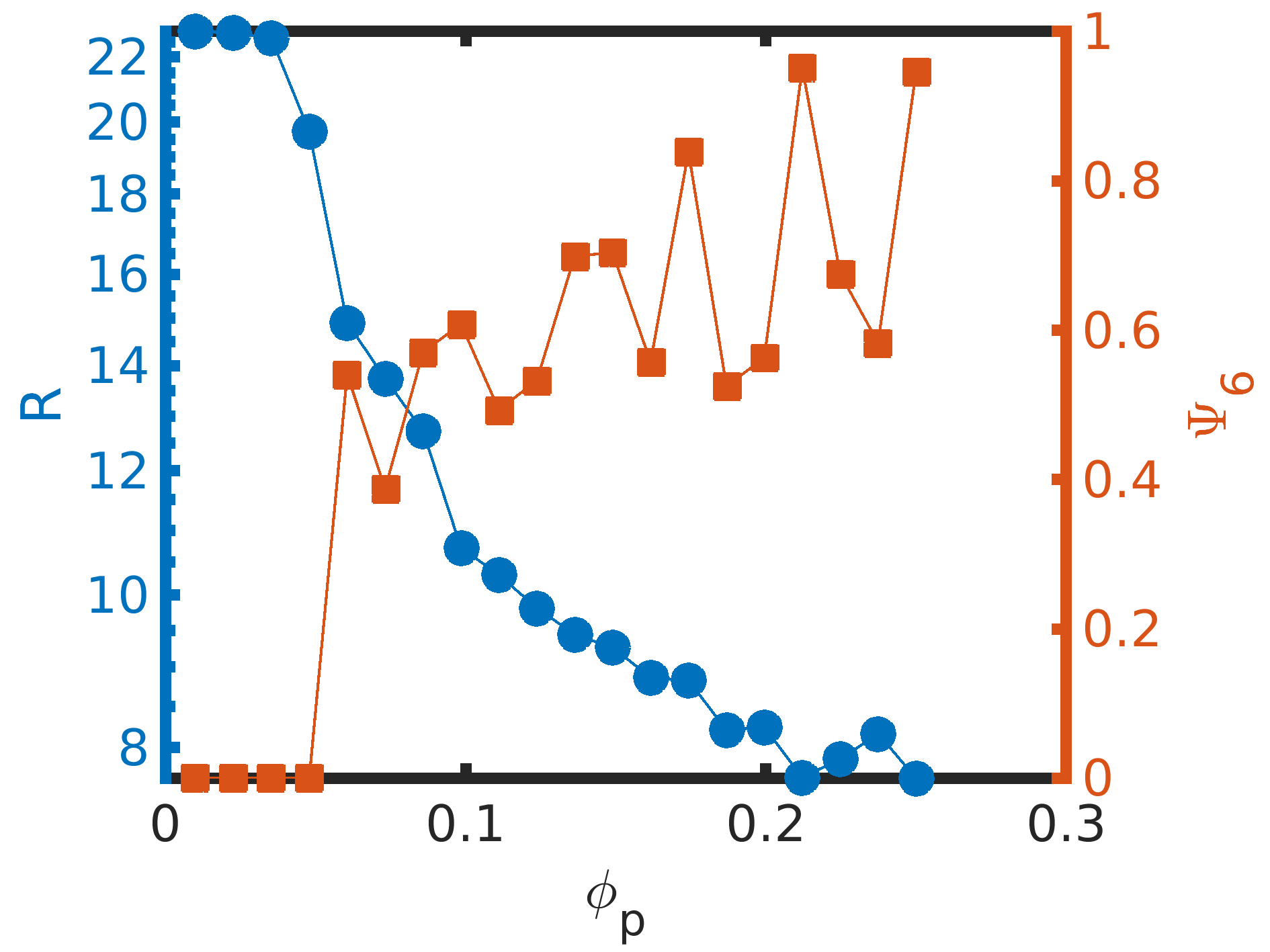}
    \caption{Activity-stabilised droplets with hexatic order in terms of AP concentration $\phi_p$ for an activity $Pe=10.1$  }
    \label{fig:concentration.droplet}
\end{figure}

In the main text we have fixed the AP concentration $\phi_p=0.2$. 
Fig. \ref{fig:concentration.droplet} studies the role of the AP concentration $\phi_p$ on the formation of regime II of stublised droplets for a fixed activity rate $Pe=10.1$. 
It is clear that a critical $\phi_p^*\approx 0.05$ can be identified. 

\subsection{Role of BM composition}

\begin{figure}
    \centering
    \includegraphics[width=1.0\linewidth]{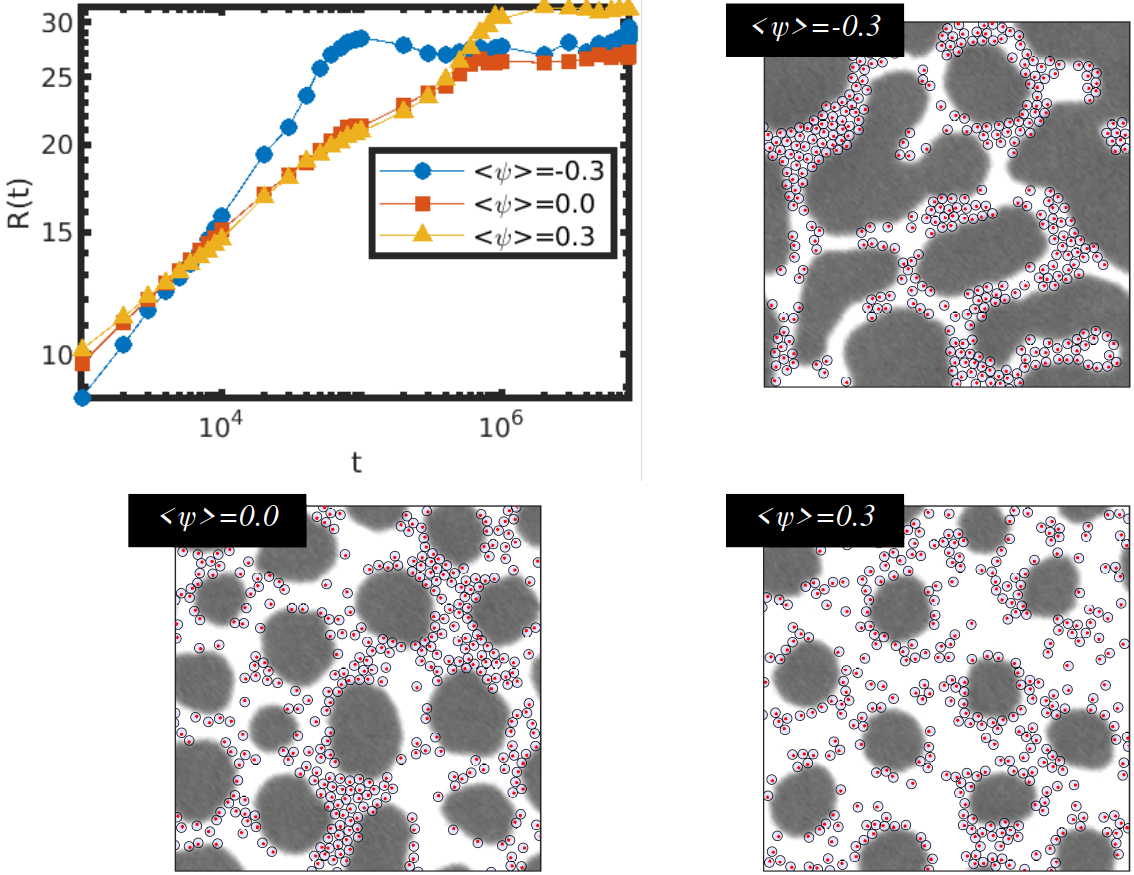}
    \caption{
    Stabilised emulsion $Pe=10.1$, $\phi_p=0.2$ for several values of the BM composition $\langle \psi \rangle$, shown as the characteristic domain size $R(t)$ over time. 
    }
    \label{fig:composition}
\end{figure}

In the main text we have assumed a symmetric BM $\langle \psi \rangle=0$. 
In Fig. \ref{fig:composition} we consider a $Pe=10.1$, $\phi_p=0.2$ regime for three values of the BM composition. 
It is clear that the arrest of the BM coarsening occurs all the values of $\langle \psi \rangle$ considered, while the formation of isotropic droplets is only favoured when the in-droplet species are not the majority phase. 

\subsection{Role of BM time scale}

\begin{figure}
    \centering
    \includegraphics[width=1.0\linewidth]{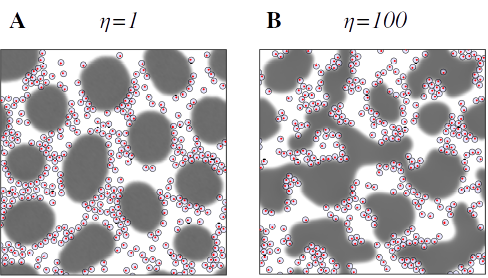}
    \caption{
    Final configuration for $Pe=10.1$ but with \textbf{A} slower BM with $\eta=1$, 
    or \textbf{B} faster BM with $\eta=100$, compared with the AP rotational time scale. 
    This is to be compared with the case of $\eta=10$ in Fig. 2. 
    }
    \label{fig:slowerBM}
\end{figure}

In the main text we have fixed the BM time scale $t_{BM}$ with respect to the passive characteristic times of the APs $t_{rot}$. 
This, in turn, specifies the relative time scale ratio $\eta=t_{rot}/t_{BM}$. 
Fig. \ref{fig:slowerBM} considers the role of $\eta$ for $Pe=10.1$. 

\subsection{Role of the initial condition: quench \textit{vs} initial equilibrium}

\begin{figure}
    \centering
    \includegraphics[width=1.0\linewidth]{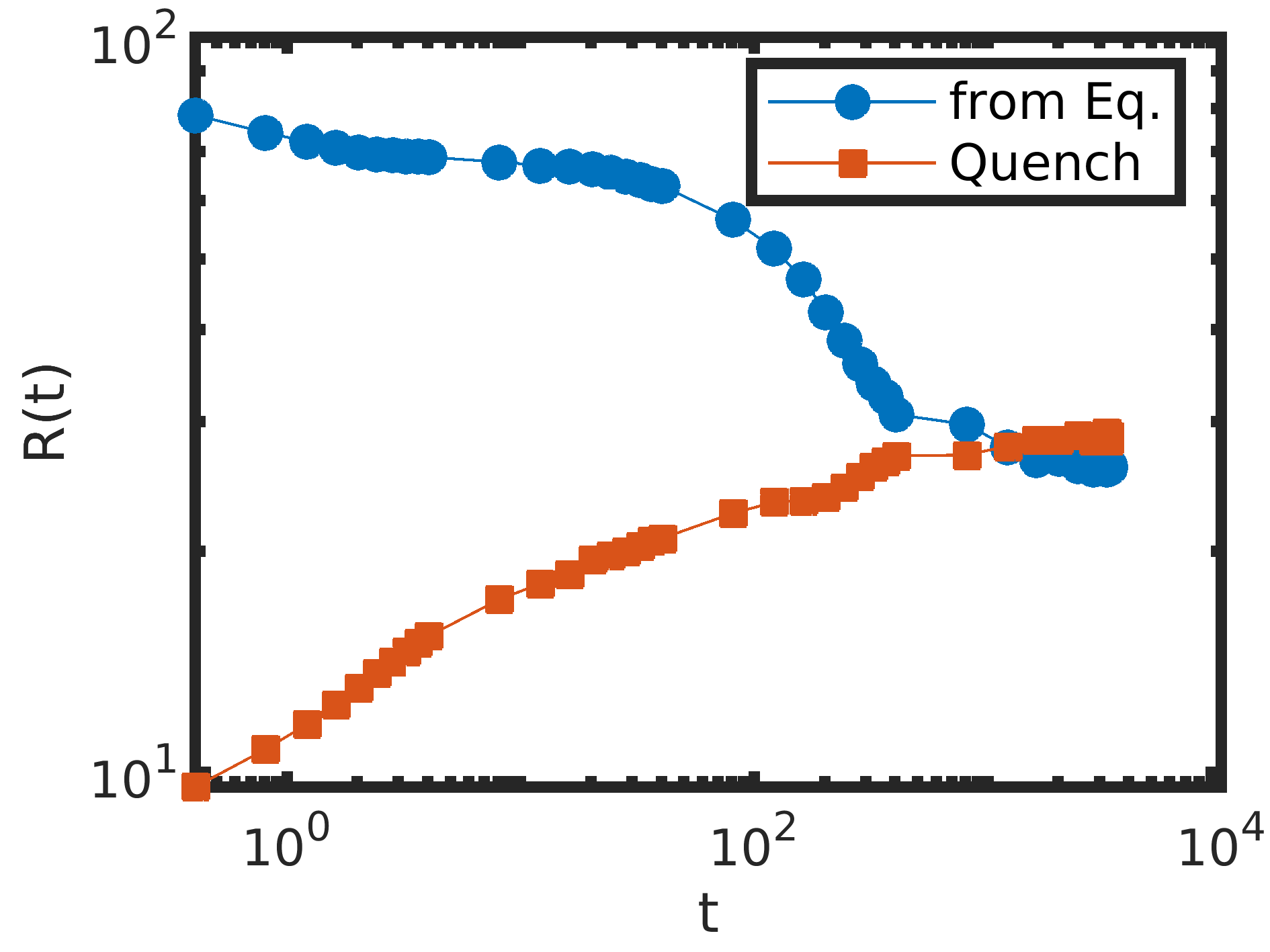}
    \caption{Comparing the emergence of hexatically-ordered droplets from a quench (activity prevents phase separation) or from equilibrium(activity drives the system away from equilibrium)}
    \label{fig:initcompare}
\end{figure}

Fig \ref{fig:initcompare} shows the characteristic droplet size in time depending on the initial configuration: 
starting from a quench by selecting a random distribution of BM and APs, and 
from equilibrium by initialising a already phase separated system. 
All other parameters are equal as Fig. \ref{fig:Rtsnaps} for $Pe=10.1$. 

\subsection{Additional snapshots of the activity-induced emulsion}

\begin{figure*}
    \centering
    \includegraphics[width=0.9\textwidth]{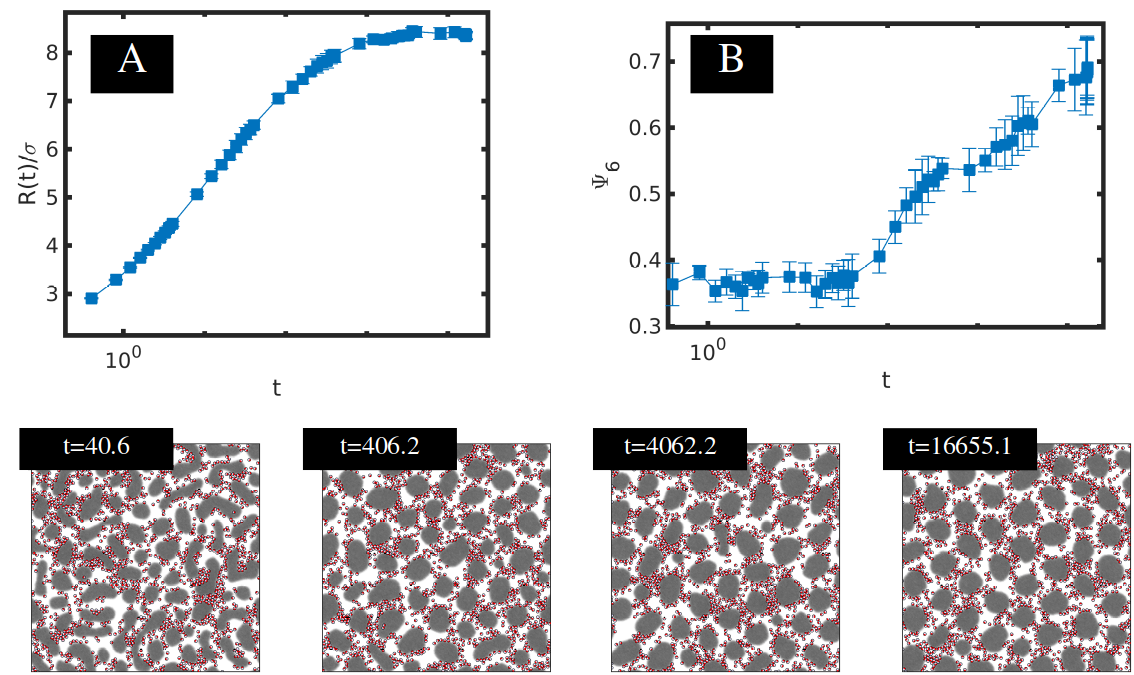}
    \caption{
    \textbf{Time evolution of the stabilised droplet phase} with $Pe=10.1$ and $\phi_p=0.2$. 
    \textbf{A} and \textbf{B} show the curves of $R(t)$ and $\Psi_6(t)$, respectively, averaged over $5$ independent runs and with error bars indicating the standard deviation
    }
    \label{fig:L256}
\end{figure*}

Fig. \ref{fig:L256} shows a large scale simulation with averages over $5$ independent runs to get insight over the variability. 

\subsection{Finite size effects}\label{sec:finite}

\begin{figure}
    \centering
    \includegraphics[width=1.0\linewidth]{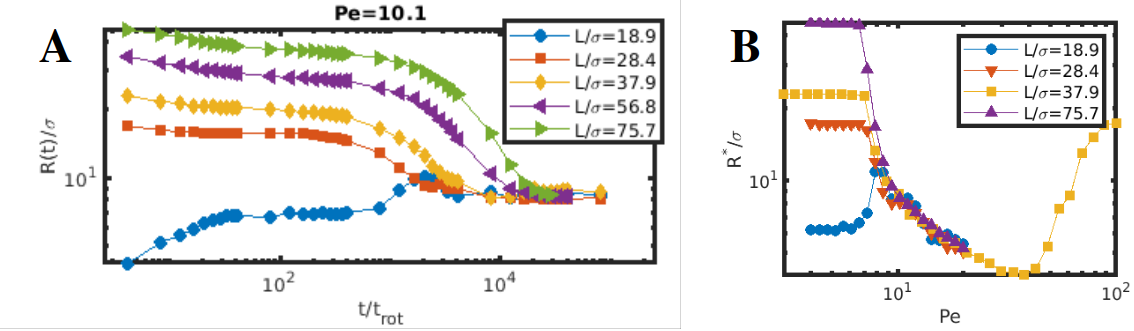}
    \caption{Finite size effects for $Pe=10.1$ for a system in a box with size $V=L^2$. 
    In \textbf{A}}
    \label{fig:finistesize}
\end{figure}

Firstly, Fig. \ref{fig:finistesize} \textbf{A} shows the time evolution of several systems with various box sizes for $Pe=10.1$. 
It is clear that the box size plays no role in the steady state behaviour of the activity-induced emulsification of the BM. 

Secondly, in Fig. \ref{fig:finistesize} \textbf{B} we repeat the study of the steady state droplet size $R^*$, as in Fig. 2 \textbf{B} in the main text, for various box sizes. 
It is clear that for small $Pe$ within regime I, the characteristic domain size scales with the system size, as expected. 
This a box size effect, which is easily explainable as the system is in equilibrium and it reaches macrophase separation. 
In the next section we dedicate a discussion on the thermodynamic limit as we take the limit of $V=L^2\to \infty$. 
Importantly, within regime II the droplet size is irrespective of the box size, indicating that there is no finite size effect in droplet morphology. 
 
\section{Analytical calculations}

\subsection{Stability of the droplet morphology}

We focus on the hexagonally-ordered stabilised droplets. 
We can consider $N$ droplets with average radius $R^*$.
Since the BM is symmetric,
\begin{equation}
    \langle \psi \rangle=0 \rightarrow N = \frac{V}{2\pi (R^*)^2}
\end{equation}
where $V$ is the system size, from which we can estimate an average interface length $\Gamma = 2\pi R^* N$. 

In the stable droplet phase, we can perform a force balance analysis.  
We identify two forces competing with each other: 
\begin{itemize}
    \item force active outwards $f_{\gamma}\sim 2\pi R^* \gamma_{AB}$ due to surface tension.
    \item (active) force acting inwards $f_{a}^1\sim v_a \gamma_t N_p^1$ where $N_p^1$ indicates number of particles acting on a droplet which we estimate as $N_p^1\sim N_p/N=8\phi_p(R^*/\sigma)^2$
\end{itemize}
if balance these two contributions, we get the following expression
\begin{equation}
    R/\sigma = \frac{\gamma_{AB} / (k_BT/\sigma^2)}{12 Pe \phi_p}
\end{equation}
which describes the characteristic size of a droplet stabilised by the presence of APs due to activity following the force balance estimation. 
In figure 2 \textbf{B} (yellow curve) we can see that this expression roughly predicts the behaviour of $R^*$. 
In particular, we focus on the regime II, where  hexagonal droplets emerge, which we can identify as the region where force balance is strictly satisfied. 
This region roughly spans $9<Pe<18$ and we can see is where the trend $R^*\propto Pe^{-1}$ is well satisfied.

\subsection{Discussion of the thermodynamic limit}\label{sec:thermo}
In the main text we consider a finite system with area $V$ and we performed numerical calculations. 
Here, we investigate the limit of $V\to \infty$ to assert the role of system size from an analytical perspective. 
We note that Fig. \ref{fig:finistesize} constitutes an analysis of the box size from a numerical point of view and as such it is limited in its ability to scale up to infintely large systems. 
The I-II transition involves the total active energy of the system compared with the energy associated with the interface of the BM. 
Therefore, it may be particularly sensitive to system size as it involves total quantities. 
On the other hand, the II-III and III-IV transitions are of a more local nature, involving \textit{per-particle} active energy and coupling energy acting on each particle.   

We firstly consider a completely phase separated BM, with an A-rich and a B-rich domain separated by a flat interface $\Gamma$, as the system described in Fig. 2. 
The energy competition for the I-II transition (APs driving the BM away from equilibrium) is 
\begin{equation}
    \epsilon_1 = \frac{N_p \epsilon_a}{\Gamma \gamma_{AB}}
\end{equation}
where we recall $N_p$ is the number of particles, $\epsilon_a$ is the active energy carried by each particle and $\gamma_{AB}$ is the BM interfacial tension.
For APs to drive the I-II transition, it is clear that $\epsilon_1 \gg 1$. 
For a fixed concentration $\phi_p$, the number of particles in the system scales as $N_p\propto V$, while the interface length for a completely phase-separated mixture can be roughly estimated as $\Gamma \sim V^{1/2}$. 
From this, it can be concluded that $\epsilon_1 \propto V^{1/2}$ and therefore in the thermodynamic limit as $V\to \infty$, the condition for the transition is always satisfied. 
However, this fails to take into account that, as the system becomes larger, less particles would be exposed to the interface of the BM, \textit{i.e.}, the ability of APs to exert pressure on the interface decreases. 
In order to be more precise, we can consider that, as the system becomes increasingly larger, the number of particles in the vicinity of the A-B interface is $\Phi N_p$. 
These are the particles which are close enough to the BM interface to  interact with it. 
The fraction $\Phi$ can be roughly estimated based on geometric considerations and it can be shown to scale as $\Phi \propto V^{-1/2}$. 
We can re-calculate the dimensionless energy parameter responsible for the I-II transition as 
\begin{equation}
    \epsilon_1 = \frac{\Phi N_p \epsilon_a}{\Gamma \gamma_{AB}}
\end{equation}
which, based on the previous scaling arguments given for $N_p\propto V$, $\Phi \propto V^{-1/2}$ and $\Gamma\propto V^{1/2}$, is clearly an intensive quantity which does not scale with the system size. 
In conclusion, in the scenario where APs drive the system away from equilibrium, we can expect that the system size does not play a fundamental role in the ability of APs to deform the equilibrium morphology.

We can consider the complementary scenario, there droplets of size $R^*$ are formed following a quench from high temperatures. 
Assuming a symmetric BM and imposing mass conservation, we know that the number of droplets satisfies 
\begin{equation}
    N \sim V/(2\pi (R^*)^2)
\end{equation}
from which we can estimate the droplet-droplet distance as 
\begin{equation}
    l_{dd} \sim \left( \frac{V}{N} \right)^{1/2} \sim \sqrt{2\pi} R^*
\end{equation}
which can  be compared with the persistence length $l_p=\sigma Pe$ as 
\begin{equation}
    \frac{l_{dd}}{l_p} \sim \frac{\sqrt{2\pi}}{Pe} \frac{R^*}{\sigma}
\end{equation}
from which it is possible to conclude that $l_{dd}$ and $l_p$ remain in the same order of magnitude. 
This is relevant, as it indicates that APs are persistent while moving between domain interfaces and assures that there is a high probability for them to persist at a droplet interface and exert active pressure. 

We consider that these two scenarios provide support for the validity of the mechanism of coarsening arrest described in the main text. 
APs can be expected to both drive the BM away from equilibrium and also to arrest phase separation.

\section{Video descriptions}

Videos of the relevant morphologies are included in the supporting information. 
These videos provide useful information on the dynamics of the structures. 
For clarity, we list the video names and parameters in table \ref{tab:videos}.
\begin{table}
  \centering
  \begin{tabular}{|c|c|c|c|}
    \hline
    \multirow{2}{*}{\textbf{regime}} & \multirow{2}{*}{\textbf{Pe}} & \multicolumn{2}{c|}{\textbf{filename}} \\
    \cline{3-4}
     &  & \textbf{from eq} & \textbf{quench} \\
    \hline
    \\
    I & 4.9 & \texttt{ video\_eq\_I.avi } &  \texttt{ video\_dis\_I.avi } \\
    II & 10.1 & \texttt{ video\_eq\_II.avi } &  \texttt{ video\_dis\_II.avi } \\
    III & 33.7 & \texttt{ video\_eq\_IIII.avi } &  \texttt{ video\_dis\_III.avi } \\
    IV & 88.6 & \texttt{ video\_eq\_IV.avi } &  \texttt{ video\_dis\_IV.avi } \\
    \hline
  \end{tabular}
  \caption{List of videos as supporting information and the corresponding regime and value of $Pe$.}
  \label{tab:videos}
\end{table}


\begin{thebibliography}{33}%
\makeatletter
\providecommand \@ifxundefined [1]{%
 \@ifx{#1\undefined}
}%
\providecommand \@ifnum [1]{%
 \ifnum #1\expandafter \@firstoftwo
 \else \expandafter \@secondoftwo
 \fi
}%
\providecommand \@ifx [1]{%
 \ifx #1\expandafter \@firstoftwo
 \else \expandafter \@secondoftwo
 \fi
}%
\providecommand \natexlab [1]{#1}%
\providecommand \enquote  [1]{``#1''}%
\providecommand \bibnamefont  [1]{#1}%
\providecommand \bibfnamefont [1]{#1}%
\providecommand \citenamefont [1]{#1}%
\providecommand \href@noop [0]{\@secondoftwo}%
\providecommand \href [0]{\begingroup \@sanitize@url \@href}%
\providecommand \@href[1]{\@@startlink{#1}\@@href}%
\providecommand \@@href[1]{\endgroup#1\@@endlink}%
\providecommand \@sanitize@url [0]{\catcode `\\12\catcode `\$12\catcode
  `\&12\catcode `\#12\catcode `\^12\catcode `\_12\catcode `\%12\relax}%
\providecommand \@@startlink[1]{}%
\providecommand \@@endlink[0]{}%
\providecommand \url  [0]{\begingroup\@sanitize@url \@url }%
\providecommand \@url [1]{\endgroup\@href {#1}{\urlprefix }}%
\providecommand \urlprefix  [0]{URL }%
\providecommand \Eprint [0]{\href }%
\providecommand \doibase [0]{http://dx.doi.org/}%
\providecommand \selectlanguage [0]{\@gobble}%
\providecommand \bibinfo  [0]{\@secondoftwo}%
\providecommand \bibfield  [0]{\@secondoftwo}%
\providecommand \translation [1]{[#1]}%
\providecommand \BibitemOpen [0]{}%
\providecommand \bibitemStop [0]{}%
\providecommand \bibitemNoStop [0]{.\EOS\space}%
\providecommand \EOS [0]{\spacefactor3000\relax}%
\providecommand \BibitemShut  [1]{\csname bibitem#1\endcsname}%
\let\auto@bib@innerbib\@empty
\bibitem [{\citenamefont {Ferrari}\ \emph {et~al.}(2022)\citenamefont
  {Ferrari}, \citenamefont {Handgraaf}, \citenamefont {Boccardo}, \citenamefont
  {Buffo}, \citenamefont {Vanni},\ and\ \citenamefont
  {Marchisio}}]{ferrari_molecular_2022}%
  \BibitemOpen
  \bibfield  {author} {\bibinfo {author} {\bibfnamefont {M.}~\bibnamefont
  {Ferrari}}, \bibinfo {author} {\bibfnamefont {J.-W.}\ \bibnamefont
  {Handgraaf}}, \bibinfo {author} {\bibfnamefont {G.}~\bibnamefont {Boccardo}},
  \bibinfo {author} {\bibfnamefont {A.}~\bibnamefont {Buffo}}, \bibinfo
  {author} {\bibfnamefont {M.}~\bibnamefont {Vanni}}, \ and\ \bibinfo {author}
  {\bibfnamefont {D.~L.}\ \bibnamefont {Marchisio}},\ }\href {\doibase
  10.1063/5.0079883} {\bibfield  {journal} {\bibinfo  {journal} {Physics of
  Fluids}\ }\textbf {\bibinfo {volume} {34}},\ \bibinfo {pages} {021903}
  (\bibinfo {year} {2022})}\BibitemShut {NoStop}%
\bibitem [{\citenamefont {Lifshitz}\ and\ \citenamefont
  {Slyozov}(1961)}]{lifshitz_kinetics_1961}%
  \BibitemOpen
  \bibfield  {author} {\bibinfo {author} {\bibfnamefont {I.~M.}\ \bibnamefont
  {Lifshitz}}\ and\ \bibinfo {author} {\bibfnamefont {V.~V.}\ \bibnamefont
  {Slyozov}},\ }\href {\doibase 10.1016/0022-3697(61)90054-3} {\bibfield
  {journal} {\bibinfo  {journal} {Journal of Physics and Chemistry of Solids}\
  }\textbf {\bibinfo {volume} {19}},\ \bibinfo {pages} {35} (\bibinfo {year}
  {1961})}\BibitemShut {NoStop}%
\bibitem [{\citenamefont {Wagner}(1961)}]{wagner_theorie_1961}%
  \BibitemOpen
  \bibfield  {author} {\bibinfo {author} {\bibfnamefont {C.}~\bibnamefont
  {Wagner}},\ }\href {\doibase 10.1002/bbpc.19610650704} {\bibfield  {journal}
  {\bibinfo  {journal} {Zeitschrift für Elektrochemie, Berichte der
  Bunsengesellschaft für physikalische Chemie}\ }\textbf {\bibinfo {volume}
  {65}},\ \bibinfo {pages} {581} (\bibinfo {year} {1961})},\ \bibinfo {note}
  \BibitemShut
  {NoStop}%
\bibitem [{\citenamefont {König}\ \emph {et~al.}(2021)\citenamefont {König},
  \citenamefont {J. Ronsin},\ and\ \citenamefont
  {Harting}}]{konig_two-dimensional_2021}%
  \BibitemOpen
  \bibfield  {author} {\bibinfo {author} {\bibfnamefont {B.}~\bibnamefont
  {König}}, \bibinfo {author} {\bibfnamefont {O.~J.}\ \bibnamefont
  {J. Ronsin}}, \ and\ \bibinfo {author} {\bibfnamefont {J.}~\bibnamefont
  {Harting}},\ }\href {\doibase 10.1039/D1CP03229A} {\bibfield  {journal}
  {\bibinfo  {journal} {Physical Chemistry Chemical Physics}\ }\textbf
  {\bibinfo {volume} {23}},\ \bibinfo {pages} {24823} (\bibinfo {year}
  {2021})},\ \bibinfo {note} {publisher: Royal Society of
  Chemistry}\BibitemShut {NoStop}%
\bibitem [{\citenamefont {Balazs}\ \emph {et~al.}(2000)\citenamefont {Balazs},
  \citenamefont {Ginzburg}, \citenamefont {Qiu}, \citenamefont {Peng},\ and\
  \citenamefont {Jasnow}}]{balazs_multi-scale_2000}%
  \BibitemOpen
  \bibfield  {author} {\bibinfo {author} {\bibfnamefont {A.~C.}\ \bibnamefont
  {Balazs}}, \bibinfo {author} {\bibfnamefont {V.~V.}\ \bibnamefont
  {Ginzburg}}, \bibinfo {author} {\bibfnamefont {F.}~\bibnamefont {Qiu}},
  \bibinfo {author} {\bibfnamefont {G.}~\bibnamefont {Peng}}, \ and\ \bibinfo
  {author} {\bibfnamefont {D.}~\bibnamefont {Jasnow}},\ }\href@noop {} {\ ,\
  \bibinfo {pages} {12} (\bibinfo {year} {2000})}\BibitemShut {NoStop}%
\bibitem [{\citenamefont {Jansen}\ and\ \citenamefont
  {Harting}(2011)}]{jansen_bijels_2011}%
  \BibitemOpen
  \bibfield  {author} {\bibinfo {author} {\bibfnamefont {F.}~\bibnamefont
  {Jansen}}\ and\ \bibinfo {author} {\bibfnamefont {J.}~\bibnamefont
  {Harting}},\ }\href {\doibase 10.1103/PhysRevE.83.046707} {\bibfield
  {journal} {\bibinfo  {journal} {Physical Review E}\ }\textbf {\bibinfo
  {volume} {83}},\ \bibinfo {pages} {046707} (\bibinfo {year} {2011})},\
  \bibinfo {note} {publisher: American Physical Society}\BibitemShut {NoStop}%
\bibitem [{\citenamefont {Stratford}\ \emph {et~al.}(2005)\citenamefont
  {Stratford}, \citenamefont {Adhikari}, \citenamefont {Pagonabarraga},
  \citenamefont {Desplat},\ and\ \citenamefont
  {Cates}}]{stratford_colloidal_2005}%
  \BibitemOpen
  \bibfield  {author} {\bibinfo {author} {\bibfnamefont {K.}~\bibnamefont
  {Stratford}}, \bibinfo {author} {\bibfnamefont {R.}~\bibnamefont {Adhikari}},
  \bibinfo {author} {\bibfnamefont {I.}~\bibnamefont {Pagonabarraga}}, \bibinfo
  {author} {\bibfnamefont {J.-C.}\ \bibnamefont {Desplat}}, \ and\ \bibinfo
  {author} {\bibfnamefont {M.~E.}\ \bibnamefont {Cates}},\ }\href {\doibase
  10.1126/science.1116589} {\bibfield  {journal} {\bibinfo  {journal}
  {Science}\ }\textbf {\bibinfo {volume} {309}},\ \bibinfo {pages} {2198}
  (\bibinfo {year} {2005})}\BibitemShut {NoStop}%
\bibitem [{\citenamefont {Gonnella}\ \emph {et~al.}(2015)\citenamefont
  {Gonnella}, \citenamefont {Marenduzzo}, \citenamefont {Suma},\ and\
  \citenamefont {Tiribocchi}}]{gonnella_motility-induced_2015}%
  \BibitemOpen
  \bibfield  {author} {\bibinfo {author} {\bibfnamefont {G.}~\bibnamefont
  {Gonnella}}, \bibinfo {author} {\bibfnamefont {D.}~\bibnamefont
  {Marenduzzo}}, \bibinfo {author} {\bibfnamefont {A.}~\bibnamefont {Suma}}, \
  and\ \bibinfo {author} {\bibfnamefont {A.}~\bibnamefont {Tiribocchi}},\
  }\href {\doibase 10.1016/j.crhy.2015.05.001} {\bibfield  {journal} {\bibinfo
  {journal} {Comptes Rendus Physique}\ }\bibinfo {series} {Coarsening dynamics
  / {Dynamique} de coarsening},\ \textbf {\bibinfo {volume} {16}},\ \bibinfo
  {pages} {316} (\bibinfo {year} {2015})}\BibitemShut {NoStop}%
\bibitem [{\citenamefont {Caporusso}\ \emph {et~al.}(2022)\citenamefont
  {Caporusso}, \citenamefont {Cugliandolo}, \citenamefont {Digregorio},
  \citenamefont {Gonnella}, \citenamefont {Levis},\ and\ \citenamefont
  {Suma}}]{caporusso_dynamics_2022}%
  \BibitemOpen
  \bibfield  {author} {\bibinfo {author} {\bibfnamefont {C.~B.}\ \bibnamefont
  {Caporusso}}, \bibinfo {author} {\bibfnamefont {L.~F.}\ \bibnamefont
  {Cugliandolo}}, \bibinfo {author} {\bibfnamefont {P.}~\bibnamefont
  {Digregorio}}, \bibinfo {author} {\bibfnamefont {G.}~\bibnamefont
  {Gonnella}}, \bibinfo {author} {\bibfnamefont {D.}~\bibnamefont {Levis}}, \
  and\ \bibinfo {author} {\bibfnamefont {A.}~\bibnamefont {Suma}},\ }\href
  {\doibase 10.48550/arXiv.2211.12361} {\enquote {\bibinfo {title} {Dynamics of
  {Motility}-{Induced} clusters: coarsening beyond {Ostwald} ripening},}\ }
  (\bibinfo {year} {2022}),\ \bibinfo {note} {arXiv:2211.12361
  [cond-mat]}\BibitemShut {NoStop}%
\bibitem [{\citenamefont {Tremaine}(2003)}]{tremaine_origin_2003}%
  \BibitemOpen
  \bibfield  {author} {\bibinfo {author} {\bibfnamefont {S.}~\bibnamefont
  {Tremaine}},\ }\href {\doibase 10.1086/345963} {\bibfield  {journal}
  {\bibinfo  {journal} {The Astronomical Journal}\ }\textbf {\bibinfo {volume}
  {125}},\ \bibinfo {pages} {894} (\bibinfo {year} {2003})}\BibitemShut
  {NoStop}%
\bibitem [{\citenamefont {Tailleur}\ and\ \citenamefont
  {Cates}(2008)}]{tailleur_statistical_2008}%
  \BibitemOpen
  \bibfield  {author} {\bibinfo {author} {\bibfnamefont {J.}~\bibnamefont
  {Tailleur}}\ and\ \bibinfo {author} {\bibfnamefont {M.~E.}\ \bibnamefont
  {Cates}},\ }\href {\doibase 10.1103/PhysRevLett.100.218103} {\bibfield
  {journal} {\bibinfo  {journal} {Physical Review Letters}\ }\textbf {\bibinfo
  {volume} {100}} (\bibinfo {year} {2008}),\ 10.1103/PhysRevLett.100.218103},\
  \bibinfo {note} {arXiv: 0803.1069}\BibitemShut {NoStop}%
\bibitem [{\citenamefont {Cates}\ and\ \citenamefont
  {Tailleur}(2015)}]{cates_motility-induced_2015}%
  \BibitemOpen
  \bibfield  {author} {\bibinfo {author} {\bibfnamefont {M.~E.}\ \bibnamefont
  {Cates}}\ and\ \bibinfo {author} {\bibfnamefont {J.}~\bibnamefont
  {Tailleur}},\ }\href {\doibase 10.1146/annurev-conmatphys-031214-014710}
  {\bibfield  {journal} {\bibinfo  {journal} {Annual Review of Condensed Matter
  Physics}\ }\textbf {\bibinfo {volume} {6}},\ \bibinfo {pages} {219} (\bibinfo
  {year} {2015})},\ \bibinfo {note} {\_eprint:
  https://doi.org/10.1146/annurev-conmatphys-031214-014710}\BibitemShut
  {NoStop}%
\bibitem [{\citenamefont {Digregorio}\ \emph {et~al.}(2018)\citenamefont
  {Digregorio}, \citenamefont {Levis}, \citenamefont {Suma}, \citenamefont
  {Cugliandolo}, \citenamefont {Gonnella},\ and\ \citenamefont
  {Pagonabarraga}}]{digregorio_full_2018}%
  \BibitemOpen
  \bibfield  {author} {\bibinfo {author} {\bibfnamefont {P.}~\bibnamefont
  {Digregorio}}, \bibinfo {author} {\bibfnamefont {D.}~\bibnamefont {Levis}},
  \bibinfo {author} {\bibfnamefont {A.}~\bibnamefont {Suma}}, \bibinfo {author}
  {\bibfnamefont {L.~F.}\ \bibnamefont {Cugliandolo}}, \bibinfo {author}
  {\bibfnamefont {G.}~\bibnamefont {Gonnella}}, \ and\ \bibinfo {author}
  {\bibfnamefont {I.}~\bibnamefont {Pagonabarraga}},\ }\href {\doibase
  10.1103/PhysRevLett.121.098003} {\bibfield  {journal} {\bibinfo  {journal}
  {Physical Review Letters}\ }\textbf {\bibinfo {volume} {121}},\ \bibinfo
  {pages} {098003} (\bibinfo {year} {2018})},\ \bibinfo {note} {publisher:
  American Physical Society}\BibitemShut {NoStop}%
\bibitem [{\citenamefont {Hagen}\ \emph {et~al.}(2011)\citenamefont {Hagen},
  \citenamefont {van Teeffelen},\ and\ \citenamefont
  {Löwen}}]{hagen_brownian_2011}%
  \BibitemOpen
  \bibfield  {author} {\bibinfo {author} {\bibfnamefont {B.~t.}\ \bibnamefont
  {Hagen}}, \bibinfo {author} {\bibfnamefont {S.}~\bibnamefont {van
  Teeffelen}}, \ and\ \bibinfo {author} {\bibfnamefont {H.}~\bibnamefont
  {Löwen}},\ }\href {\doibase 10.1088/0953-8984/23/19/194119} {\bibfield
  {journal} {\bibinfo  {journal} {Journal of Physics: Condensed Matter}\
  }\textbf {\bibinfo {volume} {23}},\ \bibinfo {pages} {194119} (\bibinfo
  {year} {2011})},\ \bibinfo {note} {arXiv: 1005.1343}\BibitemShut {NoStop}%
\bibitem [{\citenamefont {Vicsek}\ \emph {et~al.}(1995)\citenamefont {Vicsek},
  \citenamefont {Czirók}, \citenamefont {Ben-Jacob}, \citenamefont {Cohen},\
  and\ \citenamefont {Shochet}}]{vicsek_novel_1995}%
  \BibitemOpen
  \bibfield  {author} {\bibinfo {author} {\bibfnamefont {T.}~\bibnamefont
  {Vicsek}}, \bibinfo {author} {\bibfnamefont {A.}~\bibnamefont {Czirók}},
  \bibinfo {author} {\bibfnamefont {E.}~\bibnamefont {Ben-Jacob}}, \bibinfo
  {author} {\bibfnamefont {I.}~\bibnamefont {Cohen}}, \ and\ \bibinfo {author}
  {\bibfnamefont {O.}~\bibnamefont {Shochet}},\ }\href {\doibase
  10.1103/PhysRevLett.75.1226} {\bibfield  {journal} {\bibinfo  {journal}
  {Physical Review Letters}\ }\textbf {\bibinfo {volume} {75}},\ \bibinfo
  {pages} {1226} (\bibinfo {year} {1995})},\ \bibinfo {note} {publisher:
  American Physical Society}\BibitemShut {NoStop}%
\bibitem [{\citenamefont {Sesé-Sansa}\ \emph {et~al.}(2018)\citenamefont
  {Sesé-Sansa}, \citenamefont {Pagonabarraga},\ and\ \citenamefont
  {Levis}}]{sese-sansa_velocity_2018}%
  \BibitemOpen
  \bibfield  {author} {\bibinfo {author} {\bibfnamefont {E.}~\bibnamefont
  {Sesé-Sansa}}, \bibinfo {author} {\bibfnamefont {I.}~\bibnamefont
  {Pagonabarraga}}, \ and\ \bibinfo {author} {\bibfnamefont {D.}~\bibnamefont
  {Levis}},\ }\href {\doibase 10.1209/0295-5075/124/30004} {\bibfield
  {journal} {\bibinfo  {journal} {EPL (Europhysics Letters)}\ }\textbf
  {\bibinfo {volume} {124}},\ \bibinfo {pages} {30004} (\bibinfo {year}
  {2018})}\BibitemShut {NoStop}%
\bibitem [{\citenamefont {Wang}\ \emph {et~al.}(2015)\citenamefont {Wang},
  \citenamefont {In}, \citenamefont {Blanc}, \citenamefont {Nobili},\ and\
  \citenamefont {Stocco}}]{wang_enhanced_2015}%
  \BibitemOpen
  \bibfield  {author} {\bibinfo {author} {\bibfnamefont {X.}~\bibnamefont
  {Wang}}, \bibinfo {author} {\bibfnamefont {M.}~\bibnamefont {In}}, \bibinfo
  {author} {\bibfnamefont {C.}~\bibnamefont {Blanc}}, \bibinfo {author}
  {\bibfnamefont {M.}~\bibnamefont {Nobili}}, \ and\ \bibinfo {author}
  {\bibfnamefont {A.}~\bibnamefont {Stocco}},\ }\href {\doibase
  10.1039/C5SM01111F} {\bibfield  {journal} {\bibinfo  {journal} {Soft Matter}\
  }\textbf {\bibinfo {volume} {11}},\ \bibinfo {pages} {7376} (\bibinfo {year}
  {2015})},\ \bibinfo {note} {publisher: Royal Society of
  Chemistry}\BibitemShut {NoStop}%
\bibitem [{\citenamefont {Wang}\ \emph {et~al.}(2016)\citenamefont {Wang},
  \citenamefont {In}, \citenamefont {Blanc}, \citenamefont {Malgaretti},
  \citenamefont {Nobili},\ and\ \citenamefont {Stocco}}]{wang_wetting_2016}%
  \BibitemOpen
  \bibfield  {author} {\bibinfo {author} {\bibfnamefont {X.}~\bibnamefont
  {Wang}}, \bibinfo {author} {\bibfnamefont {M.}~\bibnamefont {In}}, \bibinfo
  {author} {\bibfnamefont {C.}~\bibnamefont {Blanc}}, \bibinfo {author}
  {\bibfnamefont {P.}~\bibnamefont {Malgaretti}}, \bibinfo {author}
  {\bibfnamefont {M.}~\bibnamefont {Nobili}}, \ and\ \bibinfo {author}
  {\bibfnamefont {A.}~\bibnamefont {Stocco}},\ }\href {\doibase
  10.1039/C6FD00025H} {\bibfield  {journal} {\bibinfo  {journal} {Faraday
  Discussions}\ }\textbf {\bibinfo {volume} {191}},\ \bibinfo {pages} {305}
  (\bibinfo {year} {2016})},\ \bibinfo {note} {publisher: The Royal Society of
  Chemistry}\BibitemShut {NoStop}%
\bibitem [{\citenamefont {Kokot}\ \emph {et~al.}(2022)\citenamefont {Kokot},
  \citenamefont {Faizi}, \citenamefont {Pradillo}, \citenamefont {Snezhko},\
  and\ \citenamefont {Vlahovska}}]{kokot_spontaneous_2022}%
  \BibitemOpen
  \bibfield  {author} {\bibinfo {author} {\bibfnamefont {G.}~\bibnamefont
  {Kokot}}, \bibinfo {author} {\bibfnamefont {H.~A.}\ \bibnamefont {Faizi}},
  \bibinfo {author} {\bibfnamefont {G.~E.}\ \bibnamefont {Pradillo}}, \bibinfo
  {author} {\bibfnamefont {A.}~\bibnamefont {Snezhko}}, \ and\ \bibinfo
  {author} {\bibfnamefont {P.~M.}\ \bibnamefont {Vlahovska}},\ }\href {\doibase
  10.1038/s42005-022-00872-9} {\bibfield  {journal} {\bibinfo  {journal}
  {Communications Physics}\ }\textbf {\bibinfo {volume} {5}},\ \bibinfo {pages}
  {1} (\bibinfo {year} {2022})},\ \bibinfo {note} {number: 1 Publisher: Nature
  Publishing Group}\BibitemShut {NoStop}%
\bibitem [{\citenamefont {Peterson}\ \emph {et~al.}(2021)\citenamefont
  {Peterson}, \citenamefont {Baskaran},\ and\ \citenamefont
  {Hagan}}]{peterson_vesicle_2021}%
  \BibitemOpen
  \bibfield  {author} {\bibinfo {author} {\bibfnamefont {M.~S.~E.}\
  \bibnamefont {Peterson}}, \bibinfo {author} {\bibfnamefont {A.}~\bibnamefont
  {Baskaran}}, \ and\ \bibinfo {author} {\bibfnamefont {M.~F.}\ \bibnamefont
  {Hagan}},\ }\href {http://arxiv.org/abs/2102.02733} {\bibfield  {journal}
  {\bibinfo  {journal} {arXiv:2102.02733 [cond-mat]}\ } (\bibinfo {year}
  {2021})},\ \bibinfo {note} {arXiv: 2102.02733}\BibitemShut {NoStop}%
\bibitem [{\citenamefont {Wioland}\ \emph {et~al.}(2013)\citenamefont
  {Wioland}, \citenamefont {Woodhouse}, \citenamefont {Dunkel}, \citenamefont
  {Kessler},\ and\ \citenamefont {Goldstein}}]{wioland_confinement_2013}%
  \BibitemOpen
  \bibfield  {author} {\bibinfo {author} {\bibfnamefont {H.}~\bibnamefont
  {Wioland}}, \bibinfo {author} {\bibfnamefont {F.~G.}\ \bibnamefont
  {Woodhouse}}, \bibinfo {author} {\bibfnamefont {J.}~\bibnamefont {Dunkel}},
  \bibinfo {author} {\bibfnamefont {J.~O.}\ \bibnamefont {Kessler}}, \ and\
  \bibinfo {author} {\bibfnamefont {R.~E.}\ \bibnamefont {Goldstein}},\ }\href
  {\doibase 10.1103/PhysRevLett.110.268102} {\bibfield  {journal} {\bibinfo
  {journal} {Physical Review Letters}\ }\textbf {\bibinfo {volume} {110}},\
  \bibinfo {pages} {268102} (\bibinfo {year} {2013})},\ \bibinfo {note}
  {publisher: American Physical Society}\BibitemShut {NoStop}%
\bibitem [{\citenamefont {Scholz}\ \emph {et~al.}(2021)\citenamefont {Scholz},
  \citenamefont {Ldov}, \citenamefont {Pöschel}, \citenamefont {Engel},\ and\
  \citenamefont {Löwen}}]{scholz_surfactants_2021}%
  \BibitemOpen
  \bibfield  {author} {\bibinfo {author} {\bibfnamefont {C.}~\bibnamefont
  {Scholz}}, \bibinfo {author} {\bibfnamefont {A.}~\bibnamefont {Ldov}},
  \bibinfo {author} {\bibfnamefont {T.}~\bibnamefont {Pöschel}}, \bibinfo
  {author} {\bibfnamefont {M.}~\bibnamefont {Engel}}, \ and\ \bibinfo {author}
  {\bibfnamefont {H.}~\bibnamefont {Löwen}},\ }\href {\doibase
  10.1126/sciadv.abf8998} {\bibfield  {journal} {\bibinfo  {journal} {Science
  Advances}\ }\textbf {\bibinfo {volume} {7}},\ \bibinfo {pages} {eabf8998}
  (\bibinfo {year} {2021})},\ \bibinfo {note} {publisher: American Association
  for the Advancement of Science Section: Research Article}\BibitemShut
  {NoStop}%
\bibitem [{\citenamefont {Gao}\ and\ \citenamefont
  {Li}(2017)}]{gao_self-driven_2017}%
  \BibitemOpen
  \bibfield  {author} {\bibinfo {author} {\bibfnamefont {T.}~\bibnamefont
  {Gao}}\ and\ \bibinfo {author} {\bibfnamefont {Z.}~\bibnamefont {Li}},\
  }\href {\doibase 10.1103/PhysRevLett.119.108002} {\bibfield  {journal}
  {\bibinfo  {journal} {Physical Review Letters}\ }\textbf {\bibinfo {volume}
  {119}},\ \bibinfo {pages} {108002} (\bibinfo {year} {2017})},\ \bibinfo
  {note} {publisher: American Physical Society}\BibitemShut {NoStop}%
\bibitem [{\citenamefont {Fernandez-Rodriguez}\ \emph
  {et~al.}(2020)\citenamefont {Fernandez-Rodriguez}, \citenamefont {Grillo},
  \citenamefont {Alvarez}, \citenamefont {Rathlef}, \citenamefont {Buttinoni},
  \citenamefont {Volpe},\ and\ \citenamefont
  {Isa}}]{fernandez-rodriguez_feedback-controlled_2020}%
  \BibitemOpen
  \bibfield  {author} {\bibinfo {author} {\bibfnamefont {M.~A.}\ \bibnamefont
  {Fernandez-Rodriguez}}, \bibinfo {author} {\bibfnamefont {F.}~\bibnamefont
  {Grillo}}, \bibinfo {author} {\bibfnamefont {L.}~\bibnamefont {Alvarez}},
  \bibinfo {author} {\bibfnamefont {M.}~\bibnamefont {Rathlef}}, \bibinfo
  {author} {\bibfnamefont {I.}~\bibnamefont {Buttinoni}}, \bibinfo {author}
  {\bibfnamefont {G.}~\bibnamefont {Volpe}}, \ and\ \bibinfo {author}
  {\bibfnamefont {L.}~\bibnamefont {Isa}},\ }\href {\doibase
  10.1038/s41467-020-17864-4} {\bibfield  {journal} {\bibinfo  {journal}
  {Nature Communications}\ }\textbf {\bibinfo {volume} {11}},\ \bibinfo {pages}
  {4223} (\bibinfo {year} {2020})},\ \bibinfo {note} {bandiera\_abtest: a
  Cc\_license\_type: cc\_by Cg\_type: Nature Research Journals Number: 1
  Primary\_atype: Research Publisher: Nature Publishing Group Subject\_term:
  Colloids;Statistical physics, thermodynamics and nonlinear dynamics
  Subject\_term\_id:
  colloids;statistical-physics-thermodynamics-and-nonlinear-dynamics}\BibitemShut
  {NoStop}%
\bibitem [{\citenamefont {Ohta}\ and\ \citenamefont
  {Kawasaki}(1986)}]{ohta_equilibrium_1986}%
  \BibitemOpen
  \bibfield  {author} {\bibinfo {author} {\bibfnamefont {T.}~\bibnamefont
  {Ohta}}\ and\ \bibinfo {author} {\bibfnamefont {K.}~\bibnamefont
  {Kawasaki}},\ }\href {\doibase 10.1021/ma00164a028} {\bibfield  {journal}
  {\bibinfo  {journal} {Macromolecules}\ }\textbf {\bibinfo {volume} {19}},\
  \bibinfo {pages} {2621} (\bibinfo {year} {1986})},\ \bibinfo {note}
  {publisher: American Chemical Society}\BibitemShut {NoStop}%
\bibitem [{\citenamefont {Cahn}\ and\ \citenamefont
  {Hilliard}(1958)}]{cahn_free_1958}%
  \BibitemOpen
  \bibfield  {author} {\bibinfo {author} {\bibfnamefont {J.~W.}\ \bibnamefont
  {Cahn}}\ and\ \bibinfo {author} {\bibfnamefont {J.~E.}\ \bibnamefont
  {Hilliard}},\ }\href {\doibase 10.1063/1.1744102} {\bibfield  {journal}
  {\bibinfo  {journal} {The Journal of Chemical Physics}\ }\textbf {\bibinfo
  {volume} {28}},\ \bibinfo {pages} {258} (\bibinfo {year} {1958})},\ \bibinfo
  {note} {publisher: American Institute of Physics}\BibitemShut {NoStop}%
\bibitem [{\citenamefont {Cahn}(1959)}]{cahn_free_1959}%
  \BibitemOpen
  \bibfield  {author} {\bibinfo {author} {\bibfnamefont {J.~W.}\ \bibnamefont
  {Cahn}},\ }\href {\doibase 10.1063/1.1730145} {\bibfield  {journal} {\bibinfo
   {journal} {The Journal of Chemical Physics}\ }\textbf {\bibinfo {volume}
  {30}},\ \bibinfo {pages} {1121} (\bibinfo {year} {1959})}\BibitemShut
  {NoStop}%
\bibitem [{\citenamefont {Cahn}\ and\ \citenamefont
  {Hilliard}(1959)}]{cahn_free_1959-1}%
  \BibitemOpen
  \bibfield  {author} {\bibinfo {author} {\bibfnamefont {J.~W.}\ \bibnamefont
  {Cahn}}\ and\ \bibinfo {author} {\bibfnamefont {J.~E.}\ \bibnamefont
  {Hilliard}},\ }\href {\doibase 10.1063/1.1730447} {\bibfield  {journal}
  {\bibinfo  {journal} {The Journal of Chemical Physics}\ }\textbf {\bibinfo
  {volume} {31}},\ \bibinfo {pages} {688} (\bibinfo {year} {1959})}\BibitemShut
  {NoStop}%
\bibitem [{\citenamefont {Cook}(1970)}]{cook_brownian_1970}%
  \BibitemOpen
  \bibfield  {author} {\bibinfo {author} {\bibfnamefont {H.~E.}\ \bibnamefont
  {Cook}},\ }\href {\doibase 10.1016/0001-6160(70)90144-6} {\bibfield
  {journal} {\bibinfo  {journal} {Acta Metallurgica}\ }\textbf {\bibinfo
  {volume} {18}},\ \bibinfo {pages} {297} (\bibinfo {year} {1970})}\BibitemShut
  {NoStop}%
\bibitem [{\citenamefont {Ball}\ and\ \citenamefont
  {Essery}(1990)}]{ball_spinodal_1990}%
  \BibitemOpen
  \bibfield  {author} {\bibinfo {author} {\bibfnamefont {R.~C.}\ \bibnamefont
  {Ball}}\ and\ \bibinfo {author} {\bibfnamefont {R.~L.~H.}\ \bibnamefont
  {Essery}},\ }\href {\doibase 10.1088/0953-8984/2/51/006} {\bibfield
  {journal} {\bibinfo  {journal} {Journal of Physics: Condensed Matter}\
  }\textbf {\bibinfo {volume} {2}},\ \bibinfo {pages} {10303} (\bibinfo {year}
  {1990})}\BibitemShut {NoStop}%
\bibitem [{\citenamefont {Oono}\ and\ \citenamefont
  {Puri}(1987)}]{oono_computationally_1987}%
  \BibitemOpen
  \bibfield  {author} {\bibinfo {author} {\bibfnamefont {Y.}~\bibnamefont
  {Oono}}\ and\ \bibinfo {author} {\bibfnamefont {S.}~\bibnamefont {Puri}},\
  }\href {\doibase 10.1103/PhysRevLett.58.836} {\bibfield  {journal} {\bibinfo
  {journal} {Physical Review Letters}\ }\textbf {\bibinfo {volume} {58}},\
  \bibinfo {pages} {836} (\bibinfo {year} {1987})}\BibitemShut {NoStop}%
\bibitem [{\citenamefont {Ginzburg}\ \emph {et~al.}(2000)\citenamefont
  {Ginzburg}, \citenamefont {Gibbons}, \citenamefont {Qiu}, \citenamefont
  {Peng},\ and\ \citenamefont {Balazs}}]{ginzburg_modeling_2000}%
  \BibitemOpen
  \bibfield  {author} {\bibinfo {author} {\bibfnamefont {V.~V.}\ \bibnamefont
  {Ginzburg}}, \bibinfo {author} {\bibfnamefont {C.}~\bibnamefont {Gibbons}},
  \bibinfo {author} {\bibfnamefont {F.}~\bibnamefont {Qiu}}, \bibinfo {author}
  {\bibfnamefont {G.}~\bibnamefont {Peng}}, \ and\ \bibinfo {author}
  {\bibfnamefont {A.~C.}\ \bibnamefont {Balazs}},\ }\href {\doibase
  10.1021/ma991065t} {\bibfield  {journal} {\bibinfo  {journal}
  {Macromolecules}\ }\textbf {\bibinfo {volume} {33}},\ \bibinfo {pages} {6140}
  (\bibinfo {year} {2000})}\BibitemShut {NoStop}%
\bibitem [{\citenamefont {Diaz}\ \emph {et~al.}(2022)\citenamefont {Diaz},
  \citenamefont {Pinna}, \citenamefont {Zvelindovsky},\ and\ \citenamefont
  {Pagonabarraga}}]{diaz_hybrid_2022}%
  \BibitemOpen
  \bibfield  {author} {\bibinfo {author} {\bibfnamefont {J.}~\bibnamefont
  {Diaz}}, \bibinfo {author} {\bibfnamefont {M.}~\bibnamefont {Pinna}},
  \bibinfo {author} {\bibfnamefont {A.~V.}\ \bibnamefont {Zvelindovsky}}, \
  and\ \bibinfo {author} {\bibfnamefont {I.}~\bibnamefont {Pagonabarraga}},\
  }\href {\doibase 10.3390/polym14091910} {\bibfield  {journal} {\bibinfo
  {journal} {Polymers}\ }\textbf {\bibinfo {volume} {14}},\ \bibinfo {pages}
  {1910} (\bibinfo {year} {2022})},\ \bibinfo {note} {number: 9 Publisher:
  Multidisciplinary Digital Publishing Institute}\BibitemShut {NoStop}%
  \bibitem{bechinger_active_2016}
Clemens Bechinger, Roberto Di~Leonardo, Hartmut Löwen, Charles Reichhardt,
  Giorgio Volpe, and Giovanni Volpe.
\newblock Active particles in complex and crowded environments.
\newblock {\em Reviews of Modern Physics}, 88(4):045006, November 2016.
\newblock Publisher: American Physical Society.



\bibitem{ohta_anomalous_1993}
Takao Ohta, Yoshihisa Enomoto, James~L. Harden, and Masao Doi.
\newblock Anomalous rheological behavior of ordered phases of block copolymers.
  1.
\newblock {\em Macromolecules}, 26(18):4928--4934, August 1993.


\bibitem{oono_study_1988}
Y.~Oono and S.~Puri.
\newblock Study of phase-separation dynamics by use of cell dynamical systems.
  {I}. {Modeling}.
\newblock {\em Physical Review A}, 38(1):434--453, July 1988.

\bibitem{pinna_large_2012}
M.~Pinna and A.~V. Zvelindovsky.
\newblock Large scale simulation of block copolymers with cell dynamics.
\newblock {\em The European Physical Journal B}, 85(6), June 2012.

\bibitem{solon_pressure_2015}
Alexandre~P. Solon, Joakim Stenhammar, Raphael Wittkowski, Mehran Kardar, Yariv
  Kafri, Michael~E. Cates, and Julien Tailleur.
\newblock Pressure and {Phase} {Equilibria} in {Interacting} {Active}
  {Brownian} {Spheres}.
\newblock {\em Physical Review Letters}, 114(19):198301, May 2015.
\newblock Publisher: American Physical Society.

\bibitem{speck_ideal_2016}
Thomas Speck and Robert~L. Jack.
\newblock Ideal bulk pressure of active {Brownian} particles.
\newblock {\em Physical Review E}, 93(6):062605, June 2016.
\newblock Publisher: American Physical Society.

\bibitem{tanaka_simulation_2000}
Hajime Tanaka and Takeaki Araki.
\newblock Simulation {Method} of {Colloidal} {Suspensions} with {Hydrodynamic}
  {Interactions}: {Fluid} {Particle} {Dynamics}.
\newblock {\em Physical Review Letters}, 85(6):1338--1341, August 2000.

\end{thebibliography}
\end{document}